\documentclass[pra,aps,showpacs,twocolumn,twoside,superscriptaddress]{revtex4}

\usepackage{amsmath,amsfonts,amssymb,caption,color,epsfig,graphics,graphicx,latexsym,mathrsfs,revsymb,theorem,url,verbatim,epstopdf}

\usepackage{hyperref}

\newtheorem{definition}{Definition}
\newtheorem{proposition}[definition]{Proposition}
\newtheorem{lemma}[definition]{Lemma}

\newtheorem{theorem}[definition]{Theorem}
\newtheorem{corollary}[definition]{Corollary}
\newtheorem{conjecture}[definition]{Conjecture}

\newtheorem{remark}[definition]{Remark}
\newtheorem{example}[definition]{Example}
\newtheorem{protocol}[definition]{Protocol}
\newtheorem{question}[definition]{Question}

\def\squareforqed{\hbox{\rlap{$\sqcap$}$\sqcup$}}
\def\qed{\ifmmode\squareforqed\else{\unskip\nobreak\hfil
\penalty50\hskip1em\null\nobreak\hfil\squareforqed
\parfillskip=0pt\finalhyphendemerits=0\endgraf}\fi}
\def\endenv{\ifmmode\;\else{\unskip\nobreak\hfil
\penalty50\hskip1em\null\nobreak\hfil\;
\parfillskip=0pt\finalhyphendemerits=0\endgraf}\fi}
\newenvironment{proof}{\noindent \textbf{{Proof.~} }}{\qed}
\def\Dbar{\leavevmode\lower.6ex\hbox to 0pt
{\hskip-.23ex\accent"16\hss}D}
\makeatletter
\def\url@leostyle{%
  \@ifundefined{selectfont}{\def\UrlFont{\sf}}{\def\UrlFont{\small\ttfamily}}}
\makeatother
\def\bcj{\begin{conjecture}}
\def\ecj{\end{conjecture}}
\def\bcr{\begin{corollary}}
\def\ecr{\end{corollary}}
\def\bd{\begin{definition}}
\def\ed{\end{definition}}
\def\bea{\begin{eqnarray}}
\def\eea{\end{eqnarray}}
\def\bem{\begin{enumerate}}
\def\eem{\end{enumerate}}
\def\bex{\begin{example}}
\def\eex{\end{example}}
\def\bptl{\begin{protocol}}
\def\eptl{\end{protocol}}
\def\bim{\begin{itemize}}
\def\eim{\end{itemize}}
\def\bl{\begin{lemma}}
\def\el{\end{lemma}}
\def\bpf{\begin{proof}}
\def\epf{\end{proof}}
\def\bpp{\begin{proposition}}
\def\epp{\end{proposition}}
\def\bqu{\begin{question}}
\def\equ{\end{question}}
\def\br{\begin{remark}}
\def\er{\end{remark}}
\def\bt{\begin{theorem}}
\def\et{\end{theorem}}

\def\btb{\begin{tabular}}
\def\etb{\end{tabular}}

\newcommand{\nc}{\newcommand}

\def\a{\alpha}

\def\d{\delta}

 \nc{\bA}{{\bf A}} \nc{\bB}{{\bf B}} \nc{\bC}{{\bf C}}
 \nc{\bD}{{\bf D}} \nc{\bE}{{\bf E}} \nc{\bF}{{\bf F}}
 \nc{\bG}{{\bf G}} \nc{\bH}{{\bf H}} \nc{\bI}{{\bf I}}
 \nc{\bJ}{{\bf J}} \nc{\bK}{{\bf K}} \nc{\bL}{{\bf L}}
 \nc{\bM}{{\bf M}} \nc{\bN}{{\bf N}} \nc{\bO}{{\bf O}}
 \nc{\bP}{{\bf P}} \nc{\bQ}{{\bf Q}} \nc{\bR}{{\bf R}}
 \nc{\bS}{{\bf S}} \nc{\bT}{{\bf T}} \nc{\bU}{{\bf U}}
 \nc{\bV}{{\bf V}} \nc{\bW}{{\bf W}} \nc{\bX}{{\bf X}}
 \nc{\bZ}{{\bf Z}}

\nc{\cA}{{\cal A}} \nc{\cB}{{\cal B}} \nc{\cC}{{\cal C}}
\nc{\cD}{{\cal D}} \nc{\cE}{{\cal E}} \nc{\cF}{{\cal F}}
\nc{\cG}{{\cal G}} \nc{\cH}{{\cal H}} \nc{\cI}{{\cal I}}
\nc{\cJ}{{\cal J}} \nc{\cK}{{\cal K}} \nc{\cL}{{\cal L}}
\nc{\cM}{{\cal M}} \nc{\cN}{{\cal N}} \nc{\cO}{{\cal O}}
\nc{\cP}{{\cal P}} \nc{\cQ}{{\cal Q}} \nc{\cR}{{\cal R}}
\nc{\cS}{{\cal S}} \nc{\cT}{{\cal T}} \nc{\cU}{{\cal U}}
\nc{\cV}{{\cal V}} \nc{\cW}{{\cal W}} \nc{\cX}{{\cal X}}
\nc{\cZ}{{\cal Z}}

\nc{\hA}{{\hat{A}}} \nc{\hB}{{\hat{B}}} \nc{\hC}{{\hat{C}}}
\nc{\hD}{{\hat{D}}} \nc{\hE}{{\hat{E}}} \nc{\hF}{{\hat{F}}}
\nc{\hG}{{\hat{G}}} \nc{\hH}{{\hat{H}}} \nc{\hI}{{\hat{I}}}
\nc{\hJ}{{\hat{J}}} \nc{\hK}{{\hat{K}}} \nc{\hL}{{\hat{L}}}
\nc{\hM}{{\hat{M}}} \nc{\hN}{{\hat{N}}} \nc{\hO}{{\hat{O}}}
\nc{\hP}{{\hat{P}}} \nc{\hR}{{\hat{R}}} \nc{\hS}{{\hat{S}}}
\nc{\hT}{{\hat{T}}} \nc{\hU}{{\hat{U}}} \nc{\hV}{{\hat{V}}}
\nc{\hW}{{\hat{W}}} \nc{\hX}{{\hat{X}}} \nc{\hZ}{{\hat{Z}}}

\nc{\hn}{{\hat{n}}}

\def\diag{\mathop{\rm diag}}

\def\lin{\mathop{\rm span}}

\def\max{\mathop{\rm max}}
\def\min{\mathop{\rm min}}

\def\rank{\mathop{\rm rank}}

\def\dg{\dagger}

\def\lra{\leftrightarrow}

\def\op{\oplus}
\def\ox{\otimes}

\def\ra{\rightarrow}

\newcommand{\bra}[1]{\langle#1|}
\newcommand{\ket}[1]{|#1\rangle}
\newcommand{\proj}[1]{| #1\rangle\!\langle #1 |}
\newcommand{\ketbra}[2]{|#1\rangle\!\langle#2|}

\newcommand{\abs}[1]{|#1|}

\def\Dbar{\leavevmode\lower.6ex\hbox to 0pt
{\hskip-.23ex\accent"16\hss}D}

\begin{document}
\title{Entanglement cost and entangling power of bipartite unitary and permutation operators}

\author{Lin Chen}
\affiliation{School of Mathematics and Systems Science, Beihang University, Beijing 100191, China}
\affiliation{International Research Institute for Multidisciplinary Science, Beihang University, Beijing 100191, China}
\author{Li Yu}\email{yupapers@sina.com}
\affiliation{National Institute of Informatics, 2-1-2 Hitotsubashi, Chiyoda-ku, Tokyo 101-8430, Japan}

\date{\today}

\pacs{03.67.Ac, 03.67.Lx, 03.65.Ud, 03.67.Mn}

\begin{abstract}
It is known that any bipartite unitary operator of Schmidt rank three is equivalent to a controlled unitary under local unitaries. We propose a standard form of such operators. Using the form we improve the upper bound for the entanglement cost to implement such operators under local operations and classical communications (LOCC), and provide a corresponding protocol. A part of our protocol is based on a recursive-control protocol which is helpful for implementing other unitary operators. We show that any bipartite permutation unitary of Schmidt rank three can be implemented using LOCC and two ebits. We give two protocols for implementing bipartite permutation unitaries of any Schmidt rank $r$, and showed that one of the protocol uses $O(r)$ ebits of entanglement and $O(r)$ bits of classical communication, while these two types of costs for the other protocol scale as $O(r\log r)$ but the actual values are smaller for all $r<1100$. Based on this we obtain upper bounds of the number of nonlocal CNOT gates needed to implement bipartite classical reversible maps using classical circuits under two different conditions. We also quantify the entangling power of bipartite permutation unitaries of Schmidt rank two and three. We show that they are respectively $1$ ebit and some value between $\log_2 9 - 16/9$ and $\log_2 3$ ebits.
\end{abstract}

\maketitle


\section{Introduction}\label{sec:intro}

The implementation of unitary operations is a key task in quantum information processing. Bipartite unitaries are a particularly important class to study,
because they are the base case for studying multipartite unitaries. Many tasks in quantum communication, games and cryptography are restricted to two parties. The evaluation of entanglement cost and/or classical resources for implementing unitary operations belong to a type of communication cost problems in quantum information theory. It has applications in the study of quantum networks and distributed quantum computation, see \cite{Akibue15,ick16} for recent progress on implementing nonlocal unitaries or isometries on multiple qubits, using shared entanglement in a network or using a limited set of basic gates.

Any bipartite unitary is the product of controlled unitaries \cite{bry02,blb05}. The controlled unitary can be implemented with local operations and classical communication (LOCC) and a maximally entangled state \cite{ygc10}. The entanglement cost scales with the logarithm of the number of terms of control. The number can be as large as the dimension of the controlling system. Bipartite unitaries of Schmidt rank not greater than three are equivalent to controlled unitaries under local unitaries \cite{cy13,cy14,cy14ap}. Every Schmidt-rank-two bipartite unitary can be implemented using one ebit and LOCC \cite{cy13}, but  the best upper bound for the entanglement cost of Schmidt-rank-three unitaries appears to depend on the dimensions of the Hilbert spaces: an upper bound on $d_A\times d_B$ system is $\log_2 \min\{d_A^2,d_B\}$ ebits for $d_A\le d_B$ \cite{cy14ap}. In this paper we show that all Schmidt-rank-three bipartite unitaries can be implemented using $\log_2\min\big{\{}d_A,d_B^2,4\lfloor d_B/2\rfloor+2\big{\}}$ ebits, where $A$ is the controlling side of the unitary. This is presented in Theorem \ref{thm:sch3} based on a standard form constructed in Eq. \eqref{lm:lm2}. We present a protocol for implementing some bipartite unitaries using multiple levels of control, and apply it to Schmidt-rank-three unitaries.

Reducing the entanglement cost for implementing nonlocal unitary gates is a key problem in computation or communication tasks on networks, because entanglement is often imperfect and costly to produce. A protocol that uses less entanglement would have less error in the implemented unitary gate, giving rise to less error in the final outcome of the computation or communication task. Some tasks may involve multipartite unitaries or non-unitary operations, and studying the entanglement cost of bipartite unitaries may help the study of the entanglement cost of those operations. The classical communication cost of the protocols in this paper is linear in the entanglement cost. Thus our protocols have less classical communication cost than the previous protocols. This is beneficial since classical communication is subject to noise and security concerns.

It is known that there is a dimension-independent upper bound for the entanglement cost of bipartite permutation unitaries with the help of a one-qubit ancilla on one side \cite{cy15}. The ancilla can be dropped from this statement at the cost of using more entanglement, since it can be prepared from another shared entangled pair of qubits. We construct a standard form of bipartite complex permutation unitaries of Schmidt rank $r$, when a ``big row'' of the unitary contains at least $r-1$ nonzero blocks.
(The big row is defined in Sec~\ref{sec:pre}.)
We further investigate the maximum number of distinct nonzero diagonal blocks of a controlled permutation unitary of Schmidt rank $r$. The above two results give upper bounds of entanglement cost for implementing the corresponding types of unitaries. This is presented in Lemmas \ref{le:persch} and \ref{lm:diagonal_blocks}. When the Schmidt rank is not greater than four, we give tighter upper bounds of entanglement cost in Lemmas \ref{lm:sch2} and \ref{lm:sch3perm}, and Corollary~\ref{cr:sr4perm}. In particular, any Schmidt-rank-three bipartite permutation unitary needs only $2$ ebits to implement. We give a protocol that implements any bipartite permutation unitary of Schmidt rank $r$ using $O(r \log r)$ ebits of entanglement and $O(r \log r)$ bits of classical communication. Then we present another protocol for the same task with the costs only scaling as $O(r)$, but the actual values are larger for all $r<1100$, as discussed below Theorem \ref{thm:permutation}. These results give upper bounds for the number of nonlocal CNOT gates for implementing a bipartite classical reversible map using a classical circuit under two different conditions (A \emph{nonlocal CNOT gate} is a CNOT gate that acts across the two parties, as opposed to acting locally on the bits within each party). The number is larger in the case that ancillas are required to be restored to the initial value, compared to the opposite case, and both results are under the assumption that the initial values of the ancillas are known. These results are an exponential improvement over the corresponding results in \cite{cy15}. An example of a Schmidt-rank-four permutation unitary is given in Sec.~\ref{subsec:permutation_example} with its entanglement cost analyzed. As a byproduct, we point out that the expression of bipartite complex permutation unitaries in \eqref{eq:ubigg} is further evidence supporting a recent conjecture on the ranks and marginals of multipartite states \cite{chl14}.

Classical reversible circuits may have lower energy cost compared to the circuits that involve erasures \cite{Bennett73}. The current paper touches upon the topic of classical reversible circuits, not only because our main result applies to it, but also we find that the design for the classical reversible circuits could provide hints for designing better quantum LOCC protocols or quantum unitary circuits.

The results so far are for the upper bound of entanglement cost for implementing bipartite unitaries. Another interesting topic is finding lower bounds for this quantity, such as the entangling power defined in \eqref{eq:K_e}. Any Schmidt-rank-$r$ unitary can have entangling power at most $\log_2 r$ ebits, see the beginning part of Sec.~\ref{ssec:ent_power_permutation}. In the case of $r=3$, it is much smaller than the upper bound in this paper when $d_A$ and $d_B$ are large. Recently, Soeda \textit{et al} \cite{stm11} proved that $1$ ebit of entanglement is needed for implementing any 2-qubit controlled unitary by LOCC when the resource state is of Schmidt rank two. Stahlke \textit{et al} \cite{sg11} proved that if the Schmidt rank of the resource state is equal to the Schmidt rank of the bipartite unitary, and the unitary can be implemented by the state using LOCC or separable operations, then the resource state has equal nonzero Schmidt coefficients. In Example~\ref{ex:ex2} we present a class of Schmidt-rank-three unitaries for which we do not know of a protocol with constant entanglement cost. In fact it is an open problem whether there is a constant upper bound for the entanglement cost of all Schmidt-rank-three bipartite unitaries.

Next, we show that the entangling power of any Schmidt-rank-two bipartite permutation unitary is exactly 1 ebit by Lemma \ref{le:sch2perm_entpower}. The counterpart of Schmidt-rank-three permutation unitary is some value between $\log_2 9 - 16/9$ and $\log_2 3$ ebits, as shown in Proposition \ref{le:sch3perm_entpower}. Again, there is a curious gap between the best known entanglement cost and the entangling power, similar to the case of general Schmidt-rank-three unitaries.

The rest of this paper is organized as follows. In Sec. \ref{sec:brief} we briefly introduce the appendix. In Sec.~\ref{sec:pre} we introduce the notations and preliminary lemmas used in the paper. In Sec.~\ref{sec:main} we present the main result on Schmidt-rank-three bipartite unitary operators. In Sec.~\ref{sec:permutation} we study bipartite complex permutation unitaries. We first present some preliminary lemmas, and then investigate the entanglement cost of bipartite permutation unitaries of Schmidt rank up to three in Sec.~\ref{subsec:permutation_small}, and study the protocol and entanglement cost for general bipartite permutation unitaries in Sec.~\ref{subsec:permutation_large}. An example is given in Sec.~\ref{subsec:permutation_example}, and the entangling power of bipartite permutation unitaries is studied in Sec.~\ref{ssec:ent_power_permutation}. Finally we conclude in Sec. \ref{sec:con}.

\section{Summary of technical results}
\label{sec:brief}

To enhance readability we briefly summarize the results of the current work and their relationships in this section. We have introduced Theorem \ref{thm:sch3} in the introduction, which reduces the entanglement cost to about half of the previous upper bound in \cite{cy14ap} for large classes of bipartite Schmidt-rank-three unitaries. To study this theorem, we introduce Lemma~\ref{lm:lm2} as a hard case among the possible forms of bipartite unitaries of Schmidt rank three. The proof of Theorem~\ref{thm:sch3} makes use of Protocols~\ref{ptl1b} and \ref{ptl_gp}, which are respectively a new two-level controlled unitary protocol, and a protocol from \cite{ygc10} for implementing unitaries with group-type expansion.

We study some basic properties of the real or complex bipartite permutation unitaries in terms of the Schmidt rank in Lemmas~\ref{le:persch} and \ref{lm:diagonal_blocks}. The results are used throughout Sec.~\ref{sec:permutation}. In Lemmas~\ref{lm:sch2} and \ref{lm:sch3perm} we investigate the structure and entanglement cost for (complex) permutation unitaries of Schmidt rank two or three. In Theorem \ref{thm:permutation} we show that any bipartite permutation unitary of Schmidt rank $r$ can be implemented using local operations with the help of $\min\{\log_2(B_{r+1})+r+\log_2 r, 8r-8\}$ ebits of entanglement and twice as many bits of classical communication, where $B_j$ is the Bell number defined before Lemma~\ref{le:combinations_partial_permutation}. The two terms in the result arise from Protocol~\ref{ptl2} and Protocol~\ref{ptl3}, respectively. This significantly improves over the result in Theorem 22 of \cite{cy15}, which states that such unitary can be implemented using LOCC with $3\times 2^r$ ebits.
In Theorem~\ref{thm:classical_bipartite}, we adapt the two methods of implementing bipartite permutation unitaries in the proof of Theorem~\ref{thm:permutation} to the decomposition of classical bipartite reversible circuits into local gates and nonlocal CNOT gates. In Proposition \ref{le:sch3perm_entpower}, we prove that the entangling power [defined in Eq.~\eqref{eq:K_e}] of bipartite permutation unitaries of Schmidt rank three is in the range of $[\log_2 9 - 16/9, \log_2 3]$ ebits.

\section{Preliminaries}\label{sec:pre}

In this section we introduce the notations and preliminary lemmas used in the paper. Let $\sigma_x,\sigma_y,\sigma_z$ be the usual $2\times 2$ Pauli matrices.
Denote the computational-basis states of the bipartite Hilbert space $\cH=\cH_A\ox\cH_B$ by $\ket{i,j},i=1,\cdots,d_A$, $j=1,\cdots,d_B$. Let
$I_A$ and $I_B$ be the identity
operators on the spaces $\cH_A$ and $\cH_B$, respectively. We also denote $I_d$ and $0_d$, respectively, as the identity and zero matrix of order $d$. The bipartite unitary gate $U$ acting on $\cH$ has \emph{Schmidt rank} $n$ if there is an expansion $U=\sum^n_{j=1}A_j \ox B_j$ where the $d_A\times d_A$ matrices $A_1,\cdots,A_n$ are linearly independent, and the $d_B\times d_B$ matrices $B_1,\cdots,B_n$ are also linearly independent. An equivalent definition named as the operator-Schmidt rank has been presented in \cite{Nielsen03,Tyson03}. The above expansion is called the \emph{Schmidt decomposition}.
We name the $A$ $(B)$ space of $U$ as the space spanned by all $A_j$ $(B_j)$ that appear in a Schmidt decomposition of $U$. It is well defined in the sense that the space is independent of the specific choice of the Schmidt decomposition.

Next, $U$ is a \textit{controlled unitary gate}, if $U$ is equivalent to $\sum^{d_A}_{j=1}\proj{j}\ox U_j$ or
$\sum^{d_B}_{j=1}V_j \ox \proj{j}$ via local unitaries. To be specific, $U$ is a controlled unitary from $A$ or $B$ side, respectively.
In particular, $U$ is controlled in the computational basis from $A$ side if $U=\sum^{d_A}_{j=1}\proj{j}\ox U_j$. Bipartite unitary gates of Schmidt rank two or three are equivalent to controlled unitaries via local unitaries \cite{cy13,cy14,cy14ap}.
We shall denote $V\op W$ as the ordinary direct sum of two matrices $V$ and $W$, and denote $V\op_B W$ as the direct sum of $V$ and $W$ from the $B$ side.
The latter is called the $B$-direct sum, and $V$ and $W$ respectively act on two subspaces $\cH_A\ox\cH'_B$ and $\cH_A\ox\cH''_B$ such that $\cH_B'\perp\cH_B''$.
A permutation matrix (or called ``permutation unitary'' or ``real permutation matrix'') is a unitary matrix containing elements $0$ and $1$ only. The partial permutation matrix is a matrix with elements being $0$ and $1$ only, satisfying that any row sum or column sum is not greater than $1$. So the partial permutation matrix may be not unitary. A bipartite controlled-permutation matrix $U$ is a permutation matrix controlled in the computational basis of one system, i.e., $U=\sum_j P_j \ox V_j$, where the projectors $P_jP_k=\d_{jk}P_j$, $V_j$ is a permutation unitary, and each $P_j \ox V_j$
is a \textit{term} of $U$.
A complex permutation matrix is a unitary matrix with exactly one nonzero element in each row and column. A ``big row'' of the $d_A d_B\times d_A d_B$ unitary matrix $U$ refers to a $d_B\times d_Ad_B$ submatrix given by $_A \bra{j}U$, for some $j\in\{1,\dots,d_A\}$. Similarly, a ``big column'' of $U$ refers to a $d_A d_B\times d_B$ submatrix given by $U\ket{j}_A$, for some $j\in\{1,\dots,d_A\}$. A ``block'' of $U$ refers to a $d_B\times d_B$ submatrix given by $_A \bra{j}U\ket{k}$, for some $j,k\in\{1,\dots,d_A\}$, and when $j=k$, the block is called a ``diagonal block.''

In all the protocols in this paper, the computational basis starts from $\ket{0}$ instead of $\ket{1}$. For an $n$-dimensional system, we respectively define the Fourier gate $F={1\over \sqrt{n}}\sum_{j,k=0}^{n-1} e^{2\pi i jk/n}\ketbra{j}{k}$, and the $Z$ gate usually as $Z=\sum_{j=0}^{n-1} e^{2\pi i j/n}\ketbra{j}{j}$ but sometimes generalizing the $\ketbra{j}{j}$ to a high-rank projector, see Protocol~\ref{ptl_ct}. The $Z$ basis is the computational basis. The $Z$-information means the information about which computational basis state that the state of the quantum system is in.

In this paper, the ``entanglement cost'' of a bipartite unitary $U$ is defined as
\bea\label{eq:def_cost}
E_c(U)=\inf_p E_c(p),
\eea
where $p$ is any one-shot exact deterministic LOCC protocol to implement $U$, and $E_c(p)$ is the amount of initial entanglement needed in the protocol. ``One-shot'' means that only one copy of the unitary is implemented, while the word ``exact'' excludes the case that some other unitary that might approximate the given unitary is implemented, and ``deterministic'' means that the unitary is implemented with no chance of failure. The Schmidt rank of initially entangled state and the dimension of ancillary space are finite in each protocol $p$, and there is no constant upper bound for these quantities. In the case that the resource entangled state is mixed, we suggest to use the entanglement of formation \cite{pv07} as the entanglement measure, although we do not discuss the mixed entangled state in this paper. If there is entanglement left after the protocol, subtraction of the latter from the cost would lead to definitions of assisted entanglement cost. It is beyond the scope of this paper.

The unit for entanglement is ``ebit.'' The entanglement contained in a maximally entangled pure state of Schmidt rank $N$ is regarded as $\log_2 N$ ebits. Also, to simplify the notation, every bit of classical communication used in a protocol is called a ``c-bit.'' If the classical message is a signal among $N$ equally possible signals, the amount of classical communication is regarded as $\log_2 N$ c-bits.

\subsection{Linear algebra}

Here we present a few preliminary results of linear algebra used throughout our paper.

\bl
\label{le:vandermonde}
Let $D$ be a diagonal unitary matrix. The following four statements are equivalent.
\\
(i) $D$ has at least three distinct eigenvalues;
\\
(ii) the identity, $D$ and $D^\dg$ are linearly independent;
\\
(iii) any unitary in the linear span of the identity and $D$ is proportional to one of them;
\\
(iv) any multiple of unitary in the linear span of the identity and $D$ is proportional to one of them.
\el
\bpf
$(i)\ra(ii)$. Let $x,y,z$ be the three distinct eigenvalues of $D$. Since $x,y,z$ all have modulus one, the matrix
$F=\left(
                   \begin{array}{ccc}
                     1 & x & x^* \\
                     1 & y & y^* \\
                     1 & z & z^* \\
                   \end{array}
                 \right)$ is the product of
the diagonal matrix $\diag(x^*,y^*,z^*)$ and a Vandermonde matrix with columns permuted, the latter has determinant $(y-x)(z-x)(z-y)$. Since $x,y,z$ are distinct, $F$ is invertible. Since $F$ is a submatrix of the matrix whose columns are the diagonal vectors of the identity, $D$ and $D^\dg$, the latter are linearly independent. We have proved $(i)\ra(ii)$.

$(i)\ra(iii)$. Let the unitary be $U=xI+yD$ where $x,y$ are complex numbers. We have $(xI+yD)(x^* I + y^* D^\dg) = I$, hence $xy^* D^\dg + x^*y D =(1-\abs{x}^2-\abs{y}^2)I$. Then $(i)\ra(iii)$ follows from $(ii)$, because of $(i)\ra(ii)$.

Finally the relations $(ii)\ra(i)$, $(iii)\ra(i)$ and $(iii)\lra(iv)$ are trivial. This completes the proof.
\epf

\smallskip
In the following lemma, a matrix $A$ is said to be ``block diagonal'' iff there is a permutation matrix $P$ such that $PAP^\dag=\left(
                   \begin{array}{cc}
                     A_1 & 0 \\
                     0 & A_2 \\
                     \end{array}
                 \right)$, where $A_1$ and $A_2$ are square matrices. We regard a $k\times k$ matrix as being of order $k$.

\bl
\label{le:linearcom}
Suppose $U$ is a unitary matrix of order at least two, and there is a nonzero diagonal matrix $D$ such that there is a nontrivial linear combination of $D$ and
$\tilde U=\left(
                   \begin{array}{cc}
                     0 & U \\
                     U^\dg & 0 \\
                     \end{array}
                 \right)$ that is unitary, and we denote it as $V$. Then $X^\dag V X$ is block diagonal, where $X=\left(
                   \begin{array}{cc}
                     W & 0 \\
                     0 & W \\
                     \end{array}
                 \right)$, and $W$ is an $n\times n$ unitary matrix.
\el
\bpf
By assumption, for the given $n\times n$ unitary matrix $U$, where $n\ge 2$, there exists a nonzero complex number $c$ and a nonzero diagonal matrix $D$ such that  $V:=c D+\tilde U$ is proportional to a unitary matrix of order $2n$ with $n\ge2$, where $\tilde U=\left(
                   \begin{array}{cc}
                     0 & U \\
                     U^\dg & 0 \\
                     \end{array}
                 \right)$.
This $V$ differs from the $V$ in the assertion by a constant factor, hence it suffices to prove the assertion for the current $V$.
Suppose $D=\diag(x_1,x_2,\dots,x_n,y_1,y_2,\dots,y_n)$, and the matrix elements of $U$ are $(U)_{ij}=u_{ij}$, $i,j\in\{1,\dots,n\}$. The rows of $V$ are mutually orthogonal. From that the $j$'th and $(n+k)$'th rows of $V$ are orthogonal, where $j,k\in\{1,\dots,n\}$, we have $x^{\ast}_j u^{\ast}_{jk} + u^{\ast}_{jk} y_k=0$, hence
\bea\label{eq:ykxj}
y_k=-x_j^{\ast} \mbox{ if } u_{jk}\ne 0, \quad \forall j,k.
\eea
Therefore, for any $j\in\{1,\dots,n\}$, it must be that those $x_p$ ($1\le p\le n$) that are equal to $x_j$ and those $y_q$  ($1\le q\le n$) that are equal to $-x_j^{\ast}$ satisfy that their row and column coordinates determine a rectangular block in $U$ consisting of elements $u_{pq}$, and any element of $U$ outside of this block that are in the same rows or the same columns of this block must be zero. The last statement is due to the following reason: Suppose such a rectangular block contains $u_{pq}$, then an element $u_{pq'}$ where $q'$ satisfies $y_{q'}\ne -x_p^{\ast}$ is in the row labeled by $p$ and outside of the rectangular block containing $u_{pq}$; and from \eqref{eq:ykxj}, we have $u_{pq'}=0$. Now we consider two cases:

The first case is that there exist $j,k\in\{1,\dots,n\}$ such that $x_j\ne x_k$. In this case, the $U$ contains some rectangular blocks that do not overlap in the rows and columns that they occupy. Since $U$ is unitary, these rectangular blocks must be square blocks. Hence, $U$ is block-diagonal after suitable permutation matrices are multiplied before and after it. From the form of $V$, this implies that $V$ is block diagonal in the sense defined before the lemma. Thus the assertion holds with $W$ being the identity matrix $I_n$.

The second case is that $x_1=x_2=\dots=x_n$. Then it must be that $y_1=y_2=\dots=y_n=-x_1^{\ast}$, since otherwise it can be deduced from \eqref{eq:ykxj} that there would be a column of $U$ that is zero, violating that $U$ is unitary. Since $U$ is unitary, there is an $n\times n$ diagonal matrix $E$ and an $n\times n$ unitary matrix $W$ such that $U=WEW^\dag$, then
\bea
V=\left(
                   \begin{array}{cc}
                     W & 0 \\
                     0 & W \\
                     \end{array}
                 \right) \cdot
                 \left(
                   \begin{array}{cc}
                     \gamma I_n & E \\
                     E^\dg & -\gamma^{\ast} I_n \\
                     \end{array}
                 \right) \cdot
                 \left(
                   \begin{array}{cc}
                     W^\dag & 0 \\
                     0 & W^\dag \\
                     \end{array}
                 \right),
\eea
where $\gamma=x_1$. Since $E$, $E^\dag$, and $I_n$ are all diagonal, the matrix $\left(
                   \begin{array}{cc}
                     \gamma I_n & E \\
                     E^\dg & -\gamma^{\ast} I_n \\
                     \end{array}
                 \right)$ is the direct sum of $n$ $2\times 2$ matrices up to a similarity transform by a permutation matrix. The rows and columns of the $j$'th $2\times 2$ matrix correspond to
                 the $j$'th and the $(n+j)$'th rows, and the $j$'th and the $(n+j)$'th columns of the original matrix, respectively.
                  This completes the proof.
\epf

\bl
\label{le:2x2unitary}
Any real linear combination of the three matrices $I_2$, $\left(
                   \begin{array}{cc}
                     w & 0 \\
                     0 & w^* \\
                     \end{array}
                 \right)$, and $\left(
                   \begin{array}{cc}
                     0 & x \\
                     -x^* & 0 \\
                     \end{array}
                 \right)$ is proportional to a unitary matrix.
\el
\bpf
Let $V=aI_2+b\left(
                   \begin{array}{cc}
                     w & 0 \\
                     0 & w^* \\
                     \end{array}
                 \right)+c\left(
                   \begin{array}{cc}
                     0 & x \\
                     -x^* & 0 \\
                     \end{array}
                 \right)$ where $a,b,c$ are real numbers. By direct computation one can show that $V$ is proportional to a unitary matrix.
This completes the proof.
\epf

\section{Tighter upper bound for entanglement cost of implementing Schmidt-rank-3 unitaries}
\label{sec:main}

On the problem of exact implementation of bipartite nonlocal unitaries using LOCC and shared entanglement, we use or discuss the following three known protocols. (1) The two-way teleportation protocol, i.e., teleporting the system of one party to the other party, performing the unitary there, and teleporting the system back to the original party. (2) The protocol for implementing controlled unitaries in Sec.~III of \cite{ygc10}, which is briefly reviewed as Protocol~\ref{ptl_ct} below, and it will be called ``the basic controlled-unitary protocol.'' A simple extension of it is Protocol~\ref{ptl_ct_ext}, and the latter is the basis for the two-level controlled Protocols~\ref{ptl1} and \ref{ptl1b}. (3) The group-type protocol in Sec.~IV of \cite{ygc10}, which is briefly reviewed as Protocol~\ref{ptl_gp} below. Protocol~\ref{ptl1} is used in Sec.~\ref{sec:permutation}, and Protocols~\ref{ptl1b} and \ref{ptl_gp} are used in the proof of Theorem~\ref{thm:sch3} (ii).

\bptl\label{ptl_ct} (The basic controlled unitary protocol.)

{\rm
The unitary to be implemented by two parties, Alice and Bob, is
\begin{equation}
\label{eq:basic_ct}
U=\sum_{k=0}^{N-1} P_k \otimes V_k,
\end{equation}
where $P_k$ are mutually orthogonal projectors on $\cH_A$, and $V_k$ are unitary operators on ${\cal H}_B$. The $P_k$ may be of rank greater than $1$, meaning that the dimension of $\cH_A$ may be larger than $N$.

A figure for this protocol is Fig.~5 of \cite{ygc10}. This figure was originally for the case that $P_k$ are all rank-one, but with suitable interpretation of the gates in the circuit (see Sec. III C of \cite{ygc10}), it works for the general case of higher rank $P_k$. For the protocols in this section only, the $X$ gate on a $N$-dimensional Hilbert space is defined as
\begin{equation}  \label{eqn:x_minus}
  X:= \frac{1}{\sqrt{N}}\sum_{k=0}^{N-1}\ketbra{(k-1)\mod N}{k}.
\end{equation}

The steps of the protocol are as follows.

0. The two parties initially share the following entangled state on ancillary systems $a$ and $b$, which are with Alice and Bob, respectively:
\begin{equation}\label{eqn:phi}
\vert \Phi\rangle_{ab}=\frac{1}{\sqrt{N}}\sum_{k=0}^{N-1}\vert k\rangle\otimes\vert k\rangle.
\end{equation}

1. Alice performs a controlled-$X^j$ gate $\sum_{j=0}^{N-1} P_j\ox X^j$ on systems $A$ and $a$, with $A$ as the control. (The $X^j$ means $X$ to the power $j$.) Then Alice performs a measurement on $a$ in the
standard basis, and sends the result $l$ to Bob.

2. Bob applies the gate $X^l$ to $b$. This is followed by a controlled gate $\sum_{k=0}^{N-1} \ketbra{k}{k}\ox V_k$ on $b$ and $B$, with $b$ as the control. Then Bob does a Fourier gate on $b$ (defined in Sec~\ref{sec:pre}), and measures $b$ in the standard basis. The outcome $m$ is sent to Alice.

3. Alice carries out a $Z_m = Z^{-m}$ correction on $A$, where the $Z$ is defined as $Z=\sum_{j=0}^{n-1} e^{2\pi i j/N} P_j$ (c.f. Sec. III C of \cite{ygc10}), and this definition of $Z$ reduces to that in Sec~\ref{sec:pre} in the case that all $P_k$ are rank-one. This completes the protocol.

The resource consumption of the protocol is $\log_2 N$ ebits and $2\log_2 N$ c-bits.
}
\eptl

\bptl\label{ptl_ct_ext} (The extension of the basic controlled unitary protocol to the case that some projectors in \eqref{eq:basic_ct} are replaced with zero operators.)

{\rm
If the unitary to be implemented by Alice and Bob is given by \eqref{eq:basic_ct}, but only some $P_k$ are projectors, and some others are zero operators (the output is zero for any input), then the steps of Protocol~\ref{ptl_ct} can still be carried out. Note that the controlled-$X^j$ gate in step 1 and the $Z$ gate in step 3 could be defined using the same expression as before but with the $P_j$ understood as being projectors or zero operators. The protocol still uses $\log_2 N$ ebits and $2\log_2 N$ c-bits. Suppose there are $N'<N$ operators among the $\{P_k\}$ that are nonzero; then the same unitary could be carried out with only $\log_2 N'$ ebits and $2\log_2 N'$ c-bits using Protocol~\ref{ptl_ct}. Nonetheless, the less efficient protocol turns out to be useful in Protocols~\ref{ptl1} and \ref{ptl1b} below.
}
\eptl

Next, we introduce a recursive-control protocol for implementing some bipartite unitaries with LOCC and initial entanglement.

\bptl\label{ptl1} (Protocol for implementing a bipartite unitary with two levels of control --- The special case that the lower-level controlled unitaries are controlled from a fixed side.)

{\rm
The bipartite unitary to be implemented on ${\cal H}_A\otimes {\cal H}_B$ is of the following form:
\begin{equation}\label{eq:calu}
U=\sum_{k=0}^{M-1} P_k\otimes S^E_k,
\end{equation}
where ${\cal H}_A={\cal H}_C\otimes {\cal H}_D$, and ${\cal
H}_E={\cal H}_D\otimes {\cal H}_B$, and $P_k$ are orthogonal projectors on $\cH_C$, and
\begin{equation}\label{eq:sek1}
S^E_k=\sum_{j=0}^{n_k-1} U^D_{kj} \otimes Q^{(k)}_j
\end{equation}
are controlled unitaries with local unitaries $U^D_{kj}$ on $\cH_D$. The $Q^{(k)}_j$ are projectors on $\cH_B$ and are orthogonal among different $j$ for the same $k$. Let $N:=\max\{n_k: k=0,1,\dots,M-1\}$. By introducing some zero operators to the set of $Q^{(k)}_j$ and calling the new operators $\tilde Q^{(k)}_j$, we may write all $S^E_k$ using $N$ terms:
\begin{equation}\label{eq:sek2}
S^E_k=\sum_{j=0}^{N} U^D_{kj} \otimes \tilde Q^{(k)}_j,
\end{equation}
where $U^D_{kj}$ are still local unitaries and some of them are not present in Eq.~\eqref{eq:sek1}.

The idea of the protocol can be roughly summarized as follows. The higher level of the protocol is ``$k$ controls $S^E_k$,'' and the lower level is ``$j$ controls $U^D_{kj}$.'' The steps are as follows.

0. Alice and Bob share a maximally entangled state of Schmidt rank $M$ on ${\cal H}_a\otimes{\cal H}_b$, and another maximally
entangled state of Schmidt rank $N$ on ${\cal H}_q\otimes{\cal
H}_r$. The subsystems $a$ and $q$ are on Alice's side, while
$b$ and $r$ are on Bob's side.

1. They perform the first half of the basic controlled-unitary protocol (Protocol~\ref{ptl_ct}) on ${\cal H}_C$ and ${\cal H}_a\otimes{\cal
H}_b$, until the $X^l_b$ gate in the protocol is done [the $X$ is defined in Eq.~\eqref{eqn:x_minus}]. Now they share a maximally
entangled state ${1\over\sqrt M}\sum_{k=0}^{M-1} \ket{k}_C\otimes \ket{k}_b$.

2. They perform Protocol~\ref{ptl_ct_ext} to implement $S^E_k$ using their information about $k$ stored in the entangled state above, with the help of a maximally entangled state of the form $\frac{1}{\sqrt{N}}\sum_{j=0}^{N-1}\ket{j}\ox\ket{j}$. More specifically, in the lower-level protocol, every unitary gate is controlled by the $\ket{k}_C$ state on Alice's side or the $\ket{k}_b$ on Bob's side. If there are measurements not in the standard basis in the lower-level protocol, we decompose it as a unitary followed by a measurement in the standard basis, so that all measurements are in the same basis and thus need not be controlled by information about $k$.

3. They have effectively performed the $V^B_j$ gate from the protocol in Sec.~III of \cite{ygc10}, which is the $S^E_k$
gate in the higher-level of the current protocol. Next, the subsystem $b$ is measured in the Fourier basis, and
a local unitary correction, i.e., the integer powers of the generalized $Z$ gate defined in the basic controlled-unitary protocol is done on $C$. Note that $C$ is not being measured, since it is a ``data'' system and not an ancilla.

The whole protocol uses $\log_2(MN)$ ebits and $2\log_2(MN)$ c-bits. Note that in step 2, the measurement outcomes in the lower-level protocol are the same for different controlling states labelled by $k$. This is acceptable, since the Protocol~\ref{ptl_ct_ext} (used as the lower-level protocol here) works under any measurement outcome anyway.
}
\eptl

\bptl\label{ptl1b} (Protocol for implementing a bipartite unitary with two levels of control --- The general case that the lower level unitaries are controlled from different sides.)

{\rm
In Protocol~\ref{ptl1}, the lower level unitaries are all controlled from the same side (and opposite to the direction of control in the higher level, since the case of same direction is trivial in that the unitary is then a one-level controlled unitary). Here we consider a generalization: the lower-level unitaries can be controlled from different sides. Formally, the target unitary $U$ is of the following form:
\begin{equation}\label{eq:calu2}
U=\sum_{k=0}^{M-1} P_k\otimes S^E_k,
\end{equation}
where ${\cal H}_A={\cal H}_C\otimes {\cal H}_D$, and ${\cal
H}_E={\cal H}_D\otimes {\cal H}_B$, and $P_k$ are orthogonal projectors on $\cH_C$. For each $S^E_k$, there exists an integer $n_k\ge 1$, such that at least one of the following two equations hold:
\bea\label{eq:sek34}
S^E_k&=&\sum_{j=0}^{n_k-1} U^D_{kj} \ox Q^{(k)}_j,\\
\mbox{or}\quad\quad S^E_k&=&\sum_{j=0}^{n_k-1} R^{(k)}_j \ox U^B_{kj},
\eea
where $U^D_{kj}$ and $U^B_{kj}$ are local unitaries on $\cH_D$ and $\cH_B$, respectively. The $Q^{(k)}_j$ are projectors on $\cH_B$ and are orthogonal among different $j$ for the same $k$. The $R^{(k)}_j$ are projectors on $\cH_D$ and are orthogonal among different $j$ for the same $k$.
Let $N:=\max\{n_k: k=0,1,\dots,M-1\}$. By introducing some zero operators to the set of $Q^{(k)}_j$ and $R^{(k)}_j$, and calling the new operators $\tilde Q^{(k)}_j$ or $\tilde R^{(k)}_j$, we have that for each $S^E_k$, at least one of the following two equations hold:
\bea\label{eq:sek56}
S^E_k&=&\sum_{j=0}^{N} U^D_{kj} \ox \tilde Q^{(k)}_j,\\
\mbox{or}\quad\quad S^E_k&=&\sum_{j=0}^{N} \tilde R^{(k)}_j \ox U^B_{kj},
\eea
where $U^D_{kj}$ and $U^B_{kj}$ are local unitaries on $\cH_D$ and $\cH_B$, respectively, and some of them are not present in Eq.~\eqref{eq:sek1}.

The steps of the protocol are modified from Protocol~\ref{ptl1} as follows: The first two steps are the same as the Steps 0 and 1 of Protocol~\ref{ptl1}, after which both sides have a copy of the computational-basis information of the higher-level controlling state. And since the form of the overall unitary is known, each party knows whether he or she is to act as the controlling party in the lower-level protocol, depending on the higher-level controlling state. So in the modified Step 2 of the protocol, each party does what is supposed to be done locally in the lower-level controlled-unitary protocol, with each unitary gate being controlled by the local higher-level controlling state labeled by $k$, but the measurements are all in the standard basis and thus need not be controlled (if there are measurements not in the standard basis, we decompose it as a unitary followed by a measurement in the standard basis). There are two stages of classical communication (in opposite directions) in Step 2, and for each such communication stage, the party that is supposed to send classical messages does exactly the same operations as before, but the opposite party measures in the computational basis on an extra ancilla initially in the $\frac{1}{\sqrt{N}}\sum_{j=0}^{N-1}\ket{j}$ state, and sends the outcome to the other party. The choice of measuring a useful system or a \emph{dummy} ancilla introduced above is determined by the higher level controlling state labeled by $k$. However, for actual implementation, the actual measurement should be on a fixed system. This can be resolved by a controlled-swap gate controlled by $k$, which conditionally swaps the system to be measured into a fixed system before doing the measurement. The final step is similar to Step 3 of Protocol~\ref{ptl1}.

The whole protocol requires the same amount of entanglement as in Protocol~\ref{ptl1}, but generally requires more classical communication, since the correct and \emph{dummy} messages are sent in both directions simultaneously in the two stages of classical communication in Step 2, so we allow twice as much classical communication in the lower-level protocol. Thus the overall protocol uses $\log_2 (MN)$ ebits and $2\log_2 (M N^2)$ c-bits. A dummy message is the measurement outcome of a system which was originally (before the controlled-swap gate mentioned in the previous paragraph) an ancilla in a fixed initial state. Note that the dummy classical message is only dummy for some of the higher-level controlling states labeled by $k$, but is the correct message for some others. Such message, even if ``correct'', does not carry any information about the input state for the overall unitary, by the design of the basic controlled-unitary protocol. The rationale behind the above technique is as follows: The choice of which lower-level unitary is being implemented should be indistinguishable from an outside observer, since the information about the higher-level controlling state should not be leaked to the outside observer, which is necessary for implementing a unitary operation. The reason is in Theorem~1 of \cite{ygc10}, which says that implementing a unitary successfully is equivalent to that no information about the input state of the unitary is leaked to an ``environment'' system (the tensor product of the environment system and the output system of the unitary is the entire output system of the protocol).
}
\eptl

\bptl\label{ptl_gp} (Protocol for implementing a bipartite unitary given its group-type expansion.)

{\rm
This protocol is illustrated in Fig.~8 in \cite{ygc10} (except for changes in symbols in the description below), and it implements bipartite unitaries of the form
\bea\label{eq:groupform}
U=\sum_{f\in G} V_A(f) \ox W_B(f),
\eea
where the $V_A(f)$ are unitaries acting on $\cH_A$, and they form a projective unitary representation of a finite group $G$, and $W_B(f)$ are arbitrary operators acting on $\cH_B$ but they satisfy that $U$ is unitary. This protocol uses a maximally entangled resource state of Schmidt rank $\vert G\vert$ (the order of $G$). Thus the entanglement cost is $\log_2 \vert G\vert$ ebits. The classical communication cost is $2\log_2 \vert G\vert$ c-bits. For any unitary $U$, we may expand it in the form \eqref{eq:groupform} by letting $G$ be the generalized Pauli group (ignoring overall phases) $\{X^j Z^k: j,k\in [0,d_A-1]\}$ which is of order $d_A^2$, since the $d_A^2$ generalized Pauli matrices form a basis for the space of $d_A\times d_A$ matrices.

We abbreviate the steps of the protocol here. For our purposes, a good property of the protocol to be utilized for the proof of Theorem~\ref{thm:sch3} is that when $U$ is the $A$-direct sum of some unitaries, it is often the case that there is a relatively small group $G$ (by ``small'' we mean smaller than $d_A^2$) such that $U$ can be expanded in the form \eqref{eq:groupform}. This is because of the following reason: Each component in the $A$-direct sum form of $U$ is also expandable using the form \eqref{eq:groupform}; thus, its size divided by $d_B$ is the dimension of a (projective) unitary representation of the group $G$, where the representation is obtained by restricting $V_A(f)$ to the relevant subspace of $\cH_A$, for all $f\in G$. Denote the dimension of such a projective representation as $n_i$, $i=1,\dots,K$, where $K$ is the total number of components in the $A$-direct sum form of $U$. Assume that there is a group $G$ that has inequivalent irreducible projective unitary representations of sizes $n_i$, $i=1,\dots,K'$, where $K'\ge K$, and the $n_i$ with $i>K$ (in the case $K'>K$) are arbitrary positive integers (this is, of course, a big assumption and does not hold for most bipartite unitaries, but note that we may regard several blocks in an $A$-direct sum form of $U$ as one block to increase the chance that such a group $G$ exists, which is a technique used in the proof of Theorem~\ref{thm:sch3}), then we may do the following steps: Arbitrarily choose a factor system (see the definition in \cite{ygc10}) from the set of factor systems of $G$ that admit inequivalent irreducible projective unitary representations of sizes $n_i$, $i=1,\dots,K'$ (the existence of such a factor system is guaranteed by the assumption above). Then choose a projective unitary representation of $G$ that contains all inequivalent irreducible projective unitary representations belonging to this factor system. This would be a linearly independent set of matrices according to \cite[Theorem 4]{ygc10}, and they are of a simultaneous block diagonal form. We then remove some diagonal blocks from all these matrices so that the remaining blocks are of sizes $n_i$, $i=1,\dots,K$. Then the resulting matrices would be generally linearly dependent, and from the construction, the resulting set forms a (possibly overcomplete) basis for the space of matrices with the same block structure. Thus, this set of unitary matrices can be used to expand the bipartite unitary $U$ in the form of \eqref{eq:groupform}.

In our application in the proof of Theorem~\ref{thm:sch3} in this paper, we choose the type of group $G$ directly and figure out its suitable size. A different problem has been discussed in \cite{Cohen10}, which is trying to find the smallest group $G$ when the matrix of $U$ is known. However, there is some similarity: Our reason for choosing the dihedral groups as the type of group $G$ in the proof of Theorem~\ref{thm:sch3} is based on the $B$-direct-sum form of $U$ that we proved. The algorithm for choosing the group $G$ in \cite{Cohen10} also is based on finding the $A$-direct-sum form of $U$ (which corresponds to the block diagonal structure of the operators on $\cH_A$ that are used to expand $U$).
}
\eptl

The protocols with two levels of control can be generalized to protocols with multiple levels of control. Some other generalizations are possible (but not used in this paper): The lower-level operators $S^E_k$ in the target unitary of the form \eqref{eq:calu} need not be a controlled unitary, but could be unitaries with group-type expansion in Protocol~\ref{ptl_gp}, and thus the inner level of the protocol becomes Protocol~\ref{ptl_gp}.

For studying Theorem \ref{thm:sch3}, we introduce the preliminary lemma below. We note that the simplest type of Schmidt-rank-three bipartite unitaries, which are controlled unitaries with three terms, are generally not included in Lemma~\ref{lm:lm2}, due to the restrictions on the coefficients $c_{j1},c_{j2},c_{j3}$ and the matrices $T_2$ and $T_3$ below.

\bl\label{lm:lm2}
Suppose there are three linearly independent $d\times d$ unitary matrices $I_d$, $T_2$ and $T_3$, where $I_d$ is the identity matrix, and $T_2$ is diagonal, and $T_3$ is not diagonal, and $T_2,T_3$ are not simultaneously diagonalizable under a unitary similarity transform; and $K$ distinct triplets $(c_{j1},c_{j2},c_{j3})$, where $j=1,\dots,K$, $c_{j1}$ are real and nonnegative, $c_{j2}$ and $c_{j3}$ are nonzero complex numbers, such that
\bea\label{eq:u14new2}
U=\sum_{j=1}^{K} \ketbra{j}{j} \ox (c_{j1} I_d + c_{j2} T_2 + c_{j3} T_3)
\eea
is a bipartite unitary of Schmidt rank $3$ on a $K\times d$ space $\cH_A\ox\cH_B$.

Then up to local unitaries, there is a decomposition of $U$ with the following direct sum structure on $\cH_B$:
$U=\bigoplus_{k=1}^n U_k$, $I_d=\bigoplus_{k=1}^n I^{(k)}$, $T_2=\bigoplus_{k=1}^n T_2^{(k)}$ and $T_3=\bigoplus_{k=1}^n T_3^{(k)}$, satisfying that each
\bea
\label{eq:ui}
U_k=\sum_{j=1}^{K} \ketbra{j}{j} \ox (c_{j1} I^{(k)} + c_{j2} T_2^{(k)} + c_{j3} T_3^{(k)})
\eea
is a unitary on the $K\times d_{k}$ subspace $\cH_A\ox\cH_{B_k}$ with $d=\sum_{k=1}^n d_{k}$, and that $T_3^{(1)}$ is diagonal, and for each $k>1$, $T_2^{(k)}=\diag(e^{i\a_k},-e^{-i\a_k})$, $\a_k\in\mathbb{R}$; $T_3^{(k)}$ is a non-scalar $2\times2$ unitary whose non-diagonal entries are equal and positive.
\el

The proof of this lemma is given in Appendix \ref{app:{lm:lm2}}.
Lemma~\ref{lm:lm2} leads to the following result, where assertion (i) is a structure theorem for Schmidt-rank-3 bipartite unitaries. Note that the assumption of the result implies $d_A\ge 3$ and $d_B\ge 2$.
\bt\label{thm:sch3}
Assume that $U$ is a Schmidt-rank-3 bipartite unitary controlled from the A side.  Then the following assertions hold. \\
(i) Either $U$ is the $A$-direct sum of at most three unitaries of Schmidt rank at most $2$, or $U$ is locally equivalent to a $B$-direct sum of controlled unitaries of Schmidt rank at most $3$. Each of the controlled unitaries is on a $d_A\times 1$ or $d_A\times 2$ space controlled in the computational basis of $\cH_A$.\\
(ii) $U$ can be implemented by local operations and
\bea
\label{eq:log2}
\log_2\min\big{\{}d_A,d_B^2,4\lfloor d_B/2\rfloor+2\big{\}}
\eea
ebits of entanglement and
\bea
\label{eq:log2cc}
2\log_2\min\big{\{}d_A,d_B^2,\max\{12,4\lfloor d_B/2\rfloor+2\}\big{\}}
\eea
c-bits.
\et

The proof of this theorem is given in Appendix \ref{app:{thm:sch3}}.
Given that the $A$ side is the control, the result in \cite{cy14ap} gives an entanglement cost upper bound of $\log_2\min\{d_A,d_B^2\}$ ebits. This old upper bound is always not less than the new upper bound in \eqref{eq:log2}. When $d_A, d_B$ are both large and $d_A$ is about $d_B^2$, the new upper bound in \eqref{eq:log2} is about $\log_2 (2d_B)=1+\log_2 d_B$ ebits, which is about half of the old upper bound which is about $\log_2 d_B^2=2\log_2 d_B$ ebits.

\smallskip
We show two classes of examples. The first shows that for some $U$, the entanglement cost can be much less than the upper bound in Theorem~\ref{thm:sch3}(ii).
\bex\label{ex:ex1}
\rm
Consider a Schmidt-rank-three unitary $U$ of the form \eqref{eq:u_sum_T_j}. Let $\cH_B$ be of dimension $2n$, and $T_1=I_B$, $T_2=\oplus_{j=1}^n \sigma_z$,  $T_3=\oplus_{j=1}^n [\cos(t_j)\sigma_x + \sin(t_j) \sigma_y]$, where $t_j$ ($1\le j\le n$) are some different real numbers. Then $(T_2)^2=(T_3)^2=I_B$, and $T_2 T_3 = -T_3 T_2$. Actually, by conjugation using a local diagonal unitary on $\cH_B$, we can transform $T_3$ into $\oplus_{j=1}^n \sigma_x$ while keeping $T_1$ and $T_2$ unchanged. The other $T_j$ with $j>3$ are given by $T_j=\cos\theta_j T_1+i\sin\theta_j\cos\phi_j T_2+i\sin\theta_j\sin\phi_j T_3$, where $\theta_j$ and $\phi_j$ are real. The $B$ space of $U$ is spanned by a projective representation of an Abelian group of order $4$ (the Klein-four group), hence Protocol~\ref{ptl_gp} implements $U$ using 2 ebits of entanglement and LOCC. This is much less than the upper bound in Theorem~\ref{thm:sch3}(ii) when $d_A$ and $d_B$ are large.
\eex

The second class of examples is still for unitary $U$ of the form in Lemma~\ref{lm:lm2}, but with essentially different blocks in different subspaces of $\cH_B$. It suggests that there might not be an easy improvement to the upper bound in Theorem~\ref{thm:sch3}(ii) for general Schmidt-rank-three bipartite unitaries.
\bex\label{ex:ex2}
\rm
We use the notations in the proof of Lemma~\ref{lm:lm2}, but assume that the unitary is without the diagonal part, i.e. the subspace $\cH_{B_1}$ is a null space. Assume the diagonal elements of the $2\times 2$ matrices $T_2^{(k)}$ and $D_3^{(k)}$ are $s_k \sqrt{1-b^2}+b i$ and $s_k t b\sqrt{\frac{1-b}{1+b}}+t b i$, respectively, where $b\in(0,1]$ is a variable dependent on $k$, and $t$ is a positive constant less than $1$, e.g. $t=1/2$, and the sign factor $s_k$ for the real part is either $1$ or $-1$. Suppose the diagonal elements with the positive $s_k$ appear first in each $T_2^{(k)}$ and $D_3^{(k)}$, and denote such elements as $T_{2k}$ and $D_{3k}$, respectively. Then ${\rm Im}(T_{2k})$, ${\rm Im}(D_{3k})$, and ${\rm Re}(T_{2k} D^\ast_{3k})$ are $b$, $t b$, and $t b$, respectively, which is useful for checking the result below. Since $\vert D_{3k}\vert\le1$, the two off-diagonal elements of $D_3^{(k)}$ are chosen to be equal real numbers such that $D_3^{(k)}$ is unitary. Let the $(c_{j1},c_{j2},c_{j3})$ satisfy that $c_{j1}=(t y -1)/\sqrt{(1+y^2)(t y-1)^2 + t^2 y^2}$, and $c_{j2}=i c_{j1} t y /(t y -1)$, $c_{j3}=i c_{j1} y$, for $j=1,\dots,M$, where $M$ is an arbitrary positive integer, and $y=y_j>1/t$ is a real positive number independent of $k$ but dependent on $j$.  Note that $b=b_k$ is independent of $j$.
The diagonal part of Eq.~\eqref{eq:cj_tj} can be written as
\bea\label{eq:cjtj4}
&& (c_{j1})^2 + (\tilde c_{j2})^2 + (\tilde c_{j3})^2 - 2 c_{j1} \tilde c_{j2} {\rm Im}(T_{2k}) \notag\\
&& - 2 c_{j1} \tilde c_{j3} {\rm Im}(D_{3k}) + 2 \tilde c_{j2} \tilde c_{j3} {\rm Re}(T_{2k} D^\ast_{3k})=1~~~
\eea
for $k=1,2,\dots,d$. Here we have used that $c_{j1}$ is real, and $c_{j2}$ and $c_{j3}$ are pure imaginary, and that $T_2$, $T_3$ are unitary,
and we denote $\tilde c_{j2}:={\rm Im}(c_{j2})$, $\tilde c_{j3}:={\rm Im}(c_{j3})$.
It is easily verified that there are an infinite number of solutions of $y=y_j$ and $b=b_k$ for \eqref{eq:cjtj4} when $t$ is fixed, and by choosing some sufficient but finite number of them to be used in the matrix $U$, the $U$ would have Schmidt rank three. The $U$ is unitary because each $2\times 2$ block in each controlled operator on the $B$ side is unitary, and the latter follows from Lemma~\ref{le:2x2unitary} and our choice of the $T_{2k}$ and $D_{3k}$, and that $c_{j1}$ is real, and $c_{j2}$ and $c_{j3}$ are pure imaginary. The statement about the number of solutions above implies that the dimensions $d_A$ and $d_B$ are arbitrarily large, and we do not know of any simple protocol that implements this class of unitaries with a constant number of ebits and LOCC. This suggests there might not be an easy improvement to the upper bound in Theorem~\ref{thm:sch3}(ii).
\eex

\section{Entanglement cost and entangling power of bipartite permutation unitaries}\label{sec:permutation}

This section is motivated by the following question. What is the entanglement cost for Schmidt-rank-three bipartite permutation unitaries? The result in Theorem 22 of \cite{cy15} gives an upper bound of 24 ebits, with the help of a one-qubit ancilla on one side. Other motivations to study the permutation unitaries are in the first paragraph of Sec.~\ref{subsec:permutation_example}, and also in \cite{cy15}. We shall first develop some preliminary results about bipartite (complex) permutation unitaries of general Schmidt rank, and then derive the improved upper bounds for the entanglement cost for bipartite permutation unitaries of small Schmidt rank in Sec.~\ref{subsec:permutation_small}. The case of general Schmidt rank is studied in Sec.~\ref{subsec:permutation_large}. We give an example in Sec.~\ref{subsec:permutation_example}, and study the entangling power of bipartite permutation unitaries of Schmidt rank up to three in Sec.~\ref{ssec:ent_power_permutation}.

\bl
\label{le:persch}
Let $U$ be a complex bipartite permutation matrix of Schmidt rank $r$. Then the following assertions hold.
\\
(i) The nonzero blocks in any big row or big column of $U$ are linearly independent. The number of them is at least $1$ and at most $r$.
\\
(ii) Suppose a big row of $U$ contains $r$ linearly independent blocks. Then up to local complex permutation matrices the first $r$ blocks in the big row are orthogonal projectors, whose sum is the identity matrix.

A similar statement holds when all ``row'' are replaced with ``column''.
\\
(iii) Under the assumption in (ii), up to local complex permutation matrices $U$ is a complex $r$-term controlled-permutation unitary from the $B$ side. The projectors in the terms are exactly the projectors in (ii). Such unitary can be implemented using $\log_2 r$ ebits and LOCC.\\
(iv) If $U$ is a real permutation unitary, then (ii) and (iii) hold with all occurrences of the word ``complex'' removed.
\\
(v) Suppose a big row of $U$ contains $r-1$ linearly independent blocks. Then up to local complex permutation matrices the first $r-1$ blocks in the big row are orthogonal projectors, whose sum is the identity matrix.

A similar statement holds when all ``row'' are replaced with ``column''.
\\
(vi) Under the assumption in (v), assume that the projectors and their orders are respectively $P_j$ and $s_j$ for $j=1,\cdots,r-1$. Up to local complex permutation matrices, we have
\bea
\label{eq:ubigg}
U= \bigg( (Q \ox P) \op_A
\sum^n_{j=1}(Q_j \ox P_j)
\bigg) \op_B \bigg( (\op^{r-1}_{j=n+1})_B U_j \bigg)
\notag\\
\eea
where $n\in\{0\}\cup[2,r-1]$, $P$, $Q$ and $Q_j$ are all complex permutation matrices on their respective subspaces. $P$ is of size $(\sum^{n}_{j=1} s_j) \times (\sum^{n}_{j=1} s_j)$, and the pair of matrices $Q$ and $Q_j$ ($\forall j\le n$) are orthogonal in both the input and output spaces. Furthermore, $U_j$ is a complex permutation matrix of Schmidt rank at most two on the bipartite Hilbert space $\cH_A\times \lin\{\ket{s_1+\cdots+s_n+1},\cdots,\ket{d_B}\}$. The $B$ space of $U_j$ contains $P_j$.

If $n\in[2,r-2]$, then $U$ can be implemented using $\max \{2 + \log_2 n, 2+ \log_2 (r-n-1)\}$ ebits and LOCC. If $n=0$, then $U$ can be implemented using $2+ \log_2 (r-1)$ ebits and LOCC. If $n=r-1$, then $U$ can be implemented using $1+ \log_2 (r-1)$ ebits and LOCC.
\\
(vii) In (vi), if $U$ is a real permutation unitary, and $n=0$, then under local permutations, either $U$ can be written in the $n=r-1$ case of the form of \eqref{eq:ubigg}, or $U$ is a controlled-permutation unitary controlled from the $B$ side with at most $2(r-1)$ terms, thus $U$ can be implemented using $1+ \log_2 (r-1)$ ebits of entanglement.
\el

\smallskip
The proof of this lemma is given in Appendix \ref{app:{le:persch}}.
The partial transpose has been used to study the separability problem in entanglement theory \cite{peres1996,hhh96}. Recently it has been used to study the ranks of marginals of multipartite quantum states \cite{chl14}, in terms of the following conjectured inequality
\bea
\label{eq:ineq}
\rank (\sum^k_{j=1} A_j \ox B_j ) \le k \cdot \rank (\sum^k_{j=1} A_j \ox B_j^T ),
\eea
where $A_j$ (resp. $B_j$) are matrices of the same size and $T$ denotes the transpose. In previous works we have presented a few bipartite unitaries satisfying the inequality \cite{cy14,cy14ap}. One can verify that the partial transpose of the complex permutation unitaries in (ii) and \eqref{eq:ubigg} are still unitary matrices. When considered as one of the bracket expressions in the lhs or rhs of \eqref{eq:ineq}, they both satisfy \eqref{eq:ineq}. They provide further evidence supporting the conjecture. We do not know whether all bipartite permutation matrices or complex permutation matrices satisfy \eqref{eq:ineq}.

Next we describe some simple properties about the $d_B\times d_B$ blocks in bipartite permutation matrices. Let $m(r)$ denote the maximum possible number of distinct diagonal blocks in a Schmidt-rank-$r$ bipartite controlled-permutation unitary. Let $m'(r)$ denote the maximum possible number of distinct permutation matrices in the $B$-space of a Schmidt-rank-$r$ bipartite permutation unitary. Let $n(r)$ denote the maximum possible number of distinct nonzero partial permutation matrices in the $B$-space of a Schmidt-rank-$r$ bipartite permutation unitary. Using these definitions we state the following lemma.

\bl\label{lm:diagonal_blocks}
(i) $m(r)$ is equal to the maximum number of distinct permutation matrices in the linear span of $r$ arbitrary permutation matrices of the same size.\\
(ii) $m(r)=2^{r-1}$.\\
(iii) The entanglement cost of any Schmidt-rank-$r$ controlled-permutation unitary is not more than $r-1$ ebits.\\
(iv) $m'(r)$ is not greater than the maximum number of distinct permutation matrices in the linear span of $r$ arbitrary partial permutation matrices of the same size.\\
(v) $m'(r)=2^{r-1}$.\\
(vi) $n(r)=2^r-1$, and the maximum in the definition of $n(r)$ is achieved only when the bipartite permutation unitary is equivalent to a controlled unitary from the $B$ side under local permutation unitaries.
\el

The proof of this lemma is given in Appendix \ref{app:{lm:diagonal_blocks}}.
Evidently the $m'(r)$ and $n(r)$ would be unaffected if we replace $B$ by $A$ in their definition.

\subsection{Entanglement cost of bipartite permutation unitaries of Schmidt rank two or three} \label{subsec:permutation_small}

We have studied the properties of the complex bipartite permutation unitaries in terms of the Schmidt rank in Lemmas \ref{le:persch} and \ref{lm:diagonal_blocks}. In this subsection we study the bipartite permutation unitaries of Schmidt rank two or three. They are locally equivalent to controlled unitaries \cite{cy13,cy14,cy14ap}. So they can be implemented using the basic controlled-unitary protocol by directly using the controlled form, however this might require more than minimal amount of entanglement. The Lemma~\ref{lm:sch2} (i) below, together with Lemma~\ref{le:sch2perm_entpower}, imply that the entanglement cost by directly using the controlled form is minimal for the case of Schmidt rank two.

\bl\label{lm:sch2}
(i) Any Schmidt-rank-two bipartite permutation unitary is equivalent to a two-term controlled-permutation unitary under local permutation unitaries.\\
(ii) Any Schmidt-rank-two bipartite complex permutation unitary is equivalent to a two-term controlled-complex-permutation unitary under local complex permutation unitaries.
\el
\bpf
Let us prove (ii) first. Denote the complex unitary as $U$. Its standard matrix form, also denoted by $U$, is a $d_A d_B\times d_A d_B$ matrix. If there is a big row or column of $U$ containing two nonzero blocks, then the assertion follows from Lemma \ref{le:persch}(ii)(iii). It suffices to consider the case that there is exactly one nonzero block in any big row or column of $U$. Up to local permutation matrices on $\cH_A$ we may assume that $U$ is a block-diagonal complex permutation matrix, and the first two diagonal blocks $D_1,D_2$ are linearly independent. Up to a local complex permutation matrix on $\cH_B$, we may assume $D_1=I_B$. If all diagonal blocks of $U$ are proportional to $D_1$ or $D_2$, then the assertion follows. If there is a diagonal block which is not proportional to any one of $D_1,D_2$, then $D_2$ has to be diagonal and if $D_2$ has only two distinct diagonal entries, then $U$ is equivalent to a controlled complex permutation 
unitary from the $B$ side with two terms, up to local permutation unitaries. Thus we only need to consider the remaining case, i.e., that $D_2$ is diagonal and has at least three distinct diagonal entries. However in this case $D_2$ cannot be unitary by Lemma~\ref{le:vandermonde}. This completes the proof of (ii).

The proof for (i) is similar. If there is a big row or column of $U$ containing two nonzero blocks, the assertion follows from Lemma \ref{le:persch}(iv). In the remaining case, the result follows from Lemma~\ref{lm:diagonal_blocks}(ii).
\epf

Now we investigate the structure and entanglement cost for complex permutation unitaries of Schmidt rank three. In particular, the real counterpart is completely characterized in (i).

\bl\label{lm:sch3perm}
(i) Up to local permutation unitaries, any Schmidt-rank-three bipartite permutation unitary is either equivalent to a three-term or four-term controlled-permutation unitary, or equivalent to the direct sum of a product permutation unitary and a two-term controlled-permutation unitary. Therefore such unitary can be implemented using $2$ ebits and $4$ c-bits.
\\
(ii) Any Schmidt-rank-three bipartite complex permutation unitary that is not equivalent to a diagonal unitary under local permutation unitaries can be implemented using $3$ ebits and LOCC.
\\
(iii) Any diagonal Schmidt-rank-three bipartite complex permutation unitary, whose diagonal blocks contain the identity matrix and a diagonal matrix of exactly two distinct diagonal elements, can be implemented using $2$ ebits and LOCC.
\el

The proof of this lemma is given in Appendix \ref{app:{lm:sch3perm}}.
An example for ``the direct sum of a product permutation unitary and a two-term controlled-permutation unitary'' is given by the following unitary on $3\times 2$ system:
\bea
\label{eq:uketbra11}
U&=&[\ketbra{1}{1}\ox(\ketbra{1}{2}+\ketbra{2}{1})]
\notag\\
&\oplus_A&
[(\ketbra{2}{2}+\ketbra{3}{3})\ox\ketbra{1}{1}+(\ketbra{2}{3}+\ketbra{3}{2})\ox\ketbra{2}{2}].
\notag\\
\eea

\subsection{Entanglement cost of bipartite permutation unitaries of general Schmidt rank}\label{subsec:permutation_large}

The following Protocol~\ref{ptl2} implements bipartite permutation unitary $U$ of arbitrary Schmidt rank $r$. The computational basis for each system starts with $\ket{0}$. The entanglement and classical communication cost of the protocol in terms of $r$ is analyzed in Theorem~\ref{thm:permutation}. Before introducing the protocol, we define the so-called effective input and output dimensions for $U$. An example unitary illustrating these definitions is in Example~\ref{ex4} in Sec.~\ref{subsec:permutation_example}.

\bd\label{def_state_types}

\rm
(i). The effective input dimension of $A$ is the number of types of input states of $A$. A type of input states of $A$ (or ``an input type of $A$'') is a subspace of $\cH_A$ spanned by computational basis states, so that any two big columns of $U$ corresponding to two computational basis states in the subspace have the same collection of blocks in them, ignoring the positions and the relative order of the nonzero blocks in the big column.

(2). The effective output dimension of $A$ relative to an input computational basis state of $\cH_A$ is the number of nonzero blocks in the big column of $U$ corresponding to the input computational basis state of $\cH_A$. And the labels for each effective dimension for a given input computational state of $\cH_A$ is determined by the order in which the nonzero block appears in the big column. The output computational basis state of $\cH_A$ corresponding to the big row with a nonzero block in the given big column is called an output type of $A$ relative to the input of $A$, abbreviated as ``a relative output type of $A$''.

(3). The effective output dimension of $B$ is the number of output types of $B$, where an output type of $B$ is a subspace of $\cH_B$ spanned by computational basis states, so that each computational basis state in such subspace has the same combination of being in or not in the output space of the partial permutation operators in the $B$ space of $U$. It turns out that for this definition of the output type of $B$, it suffices to consider a linearly independent set of $r$ partial permutation operators in the $B$ space of $U$, which form a basis for the $B$ space of $U$, and we call such revised definition the \emph{simplified definition}. Such a basis of $r$ partial permutation operators do exist, and they can be selected from the $d_B\times d_B$ blocks in the matrix $U$. Any other partial permutation operator in the $B$ space of $U$ is a linear combination of these $r$ basis operators. Suppose the simplified definition is inequivalent to the original definition. Then there are two computational basis states in the output space $\cH_B$ so that they are simultaneously in or not in the output space of any of the $r$ basis operators, while one and only one of them is in the output space of another partial permutation operator $Q_B$ in the $B$ space of $U$. The $Q_B$ is a linear combination of the $r$ basis operators, each of which has row sums being equal between the two said output types, hence the row sums of $Q_B$ are equal between the two said output types, and we have arrived at a contradiction. Therefore, the simplified definition is equivalent to the original definition.

\ed

\bptl\label{ptl2}
(A protocol that implements a general bipartite permutation unitary $U$.)
{\rm
\begin{figure*}[ht]
\begin{center}
\includegraphics[scale=0.9]{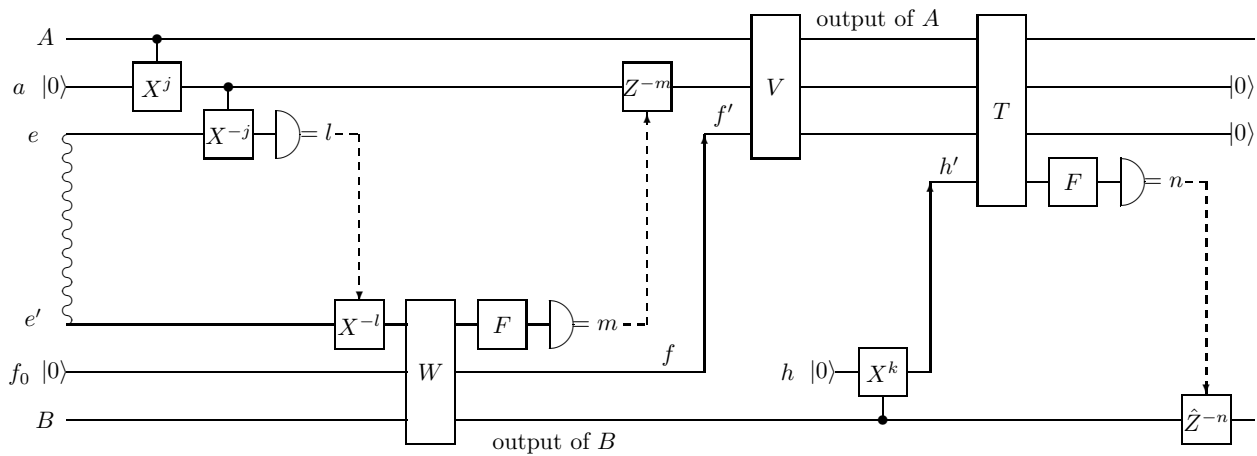}
\end{center}
\caption{The circuit diagram for Protocol~\ref{ptl2}. It implements any bipartite permutation unitary $U$ on the system $AB$, using LOCC and prior shared entanglement. The latter is explicitly shown using wavy lines or implied in the teleportation steps shown in solid vertical lines with arrows. The initial entangled state on the system $ee'$ is $\frac{1}{\sqrt{d}}\sum_k \ket{k}_e\ket{k}_{e'}$, where $d$ is the dimension of both $e$ and $e'$. The $F$ is the Fourier transform gate. The $W,V,T$ are controlled permutation gates defined in the protocol. The top input line to $W$ is in the same state as that of system $a$ after the first controlled-$X^j$ gate, which stores in its computational basis the input type of $A$. The $f$ at the second output line of $W$ is the output type of $A$ relative to the input type of $A$. The $h'$ is the output type of $B$. The $W$ is controlled by the first input line (i.e., the system $e'$), and the $V$ is controlled by the second and third lines, and the $T$ is controlled by the first and the fourth lines.} \label{fgr1}
\end{figure*}
The circuit diagram for the protocol is shown in Fig.~\ref{fgr1}. The steps of the protocol are as follows.

1. Alice prepares an ancilla $a$ in the state $\ket{0}$, and performs a controlled-$X^j$ gate on $A$ and $a$ (with projectors on $\cH_A$ of rank possibly greater than one) so that the system $a$ stores in its $Z$ basis the information about the type of input state on system $A$, which is defined in Def.~\ref{def_state_types}(i), and is abbreviated as ``the input type of $A$''. The integer $j\in\{0,1,\dots,d-1\}$ labels the type of the input state of $A$, where $d$ is the dimension of system $a$. The $X$ is the cyclic shift gate $\sum_{j=0}^{d-1} \ketbra{(j+1)\mod d}{j}$ (note it was the minus sign in \cite{ygc10} and Protocol~\ref{ptl1} instead of the plus sign).

2. Alice sends the $Z$-information about $a$ to Bob's side, so that Alice has a copy $a$ storing the $Z$-information about $a$, and Bob has a copy $e'$. This requires a prior shared maximally entangled pair of $d$-dimensional qudits $ee'$ in the state $\frac{1}{\sqrt{d}}\sum_{j=0}^{d-1} \ket{jj}$, and involves a controlled cyclic-shift gate on $ae$ and a measurement of $e$ in the standard basis on Alice's side, with the outcome sent to Bob using a classical channel, and a cyclic-shift gate on $e'$ on Bob's side according to the measurement outcome.

3. Bob has an ancilla system $f_0$ initialized in $\ket{0}$. He performs a controlled permutation unitary $W$ on $e'$ (which now stores the input type of $A$), $f_0$ and $B$, with $e'$ being the control, to prepare the output type of $A$ on the output system $f$ relative to the input $f_0$ [defined in Def.~\ref{def_state_types}(ii)], and at the same time prepare the output state of $B$ (under the action of $U$) on the system $B$. Note that if the input $f_0$ and the corresponding output $f$ for the gate $W$ are removed, the $W$ would not be unitary in general.

4. Bob measures $e'$ in the Fourier basis and a phase correction (an integer power of $Z=\sum_{j=0}^{d-1} e^{2\pi i j/d}\ketbra{j}{j}$) is done on the $a$ by Alice according to the measurement outcome sent classically. Bob teleports $f$ to the $A$ side, denoted as $f'$.

5. Alice performs a controlled permutation unitary gate $V$ on three systems $A$, $a$, and $f'$, with the joint system $af'$ being the control, to get the output of $A$.

6. The remaining task is to erase the state on $a$ and $f'$. The $a$ stores the input type of $A$, and the $f'$ stores the relative output type of $A$, and both are determined jointly by the output of $A$ together with the output type of $B$. Hence a preparation of a system $h$ containing the output type of $B$ [defined in Def.~\ref{def_state_types}(iii)] is needed, and the $h$ is teleported to the $A$ side (and denoted $h'$), for Alice to erase $a$ and $f'$ to $\ket{0}_a \ket{0}_{f'}$ by a controlled permutation unitary gate $T$ acting on $Aaf'h'$, with the joint system $Ah'$ being the control. Finally the $h'$ is measured in the Fourier basis and the outcome is sent to Bob classically, and a phase correction is done on system $B$. The phase correction gate is denoted as an integer power of $\hat Z$ to indicate that it is a diagonal operator with eigenvalues being the $d$-th roots of unity but with some degeneracies, where $d$ is the number of the output types of $B$. This completes the protocol, with the output of $U$ in systems $A$ and $B$.
\qed
}
\eptl

The following lemma gives an upper bound of the maximum number of types of the input state on system $A$ defined in Def.~\ref{def_state_types} (i). A matrix \emph{occupies a column} if and only if it has a nonzero element in that column. Suppose $S$ is a set of nonzero $d\times d$ partial permutation matrices. A subset $S'\subseteq S$ is called a \emph{covering} subset if and only if any two matrices in $S'$ do not occupy the same column, and any column is occupied by some matrix in $S'$. A \emph{basis} of $S$ is a maximal linearly independent set of matrices in $S$.

The Bell number $B_r$ is the number of different ways to partition a set of $r$ distinguishable elements, regardless of the order of partitions and the order of elements within each partition. By simple calculation, $B_1=1, B_2=2, B_3=5, B_4=15, B_5=52$, and it is known that $B_r<[0.792 r/\log_e (r+1)]^r$ for any integer $r\ge 1$ \cite{bt10}.

\bl\label{le:combinations_partial_permutation}
Suppose $S$ is a set of nonzero $d\times d$ partial permutation matrices which include exactly $r$ linearly independent matrices, and each column is occupied by some matrix in $S$. The number of covering subsets of $S$ is not greater than $B_{r+1}$.
\el
\bpf
The assertion apparently holds when $r=1$. In the following we assume $r\ge 2$. From Lemma~\ref{lm:diagonal_blocks}(vi), the size of $S$ is at most $2^r-1$. A covering subset of $S$ can contain at most $r$ elements, since elements of a covering subset must be linearly independent.

Let us fix a basis of $r$ linearly independent matrices in $S$. From the proof of Lemma~\ref{lm:diagonal_blocks}(vi), there are $r$ positions (matrix elements) of $d\times d$ matrices that determine a partial permutation matrix in the space spanned by the $r$ basis matrices. Let us call these $r$ matrix elements as ``key elements''. Some of the key elements may be in the same column. For any two matrices in the same covering subset of $S$, they occupy disjoint sets of columns, hence they cannot both contain $1$'s at the position of the same key element, nor can they contain a ``$1$'' respectively at one of two different key elements in the same column. Hence any matrix in a covering subset of $S$ is characterized by a set of key elements among the given $r$ key elements, and a covering subset of $S$ is characterized by a partitioning of the key elements, but possibly with some key element(s) not belonging to any matrix in the covering subset, in the latter case we arrange the ``extra'' key element(s) into a partition, and mark this set with an auxiliary element, i.e. let the auxiliary element and the extra key element(s) be put into the same part in the partition of $r+1$ elements.  In the case that no extra key element exists, the auxiliary element is a part of the partition by itself. Therefore, the total number of covering subsets of $S$ is at most the partition number of $r+1$ elements, which is $B_{r+1}$. This completes the proof.
\epf

\smallskip
Now we introduce a new definition of the number of input types on system $A$ (the definition for system $B$ is similar), which will be used in Protocol~\ref{ptl3} below. If the sum of all blocks in a big column of $U$ is equal to the corresponding sum for another big column, then these two big columns are regarded as \emph{of the same type in the loose sense}. The reason for this new definition is that any input computational basis state on $B$ is mapped to the same output state on $B$ under the maps represented by the two big columns which satisfy that the sum of blocks in them are equal.

\bl\label{le:input_types_A}
The number of distinct types of big columns of a bipartite permutation matrix of Schmidt rank $r$ in the loose sense is at most $2^{r-1}$. This bound is tight.
\el
\bpf
Denote the matrix as $U$, and denote the maximum value of the quantity in the assertion as $f(r)$, which is a function of $r$ only.
The sum of all blocks in a big column of $U$ is in the $B$ space of $U$, and is a matrix with elements being $0$ or $1$ and with sum of elements in each column equal to $1$. By an argument similar to that in the proof of Lemma~\ref{lm:diagonal_blocks}(ii), there are at most $2^{r-1}$ such matrices in the $B$ space of $U$. By definition, two big columns of different types in the loose sense are different in the sum of their blocks. Hence $f(r)\le 2^{r-1}$. The example of $U$ that reaches the maximum value of $2^{r-1}$ is in the proof of Lemma~\ref{lm:diagonal_blocks}(ii).
\epf

\bptl\label{ptl3}
(Another protocol that implements a general bipartite permutation unitary $U$.)
{\rm
\begin{figure*}[ht]
\begin{center}
\includegraphics[scale=0.8]{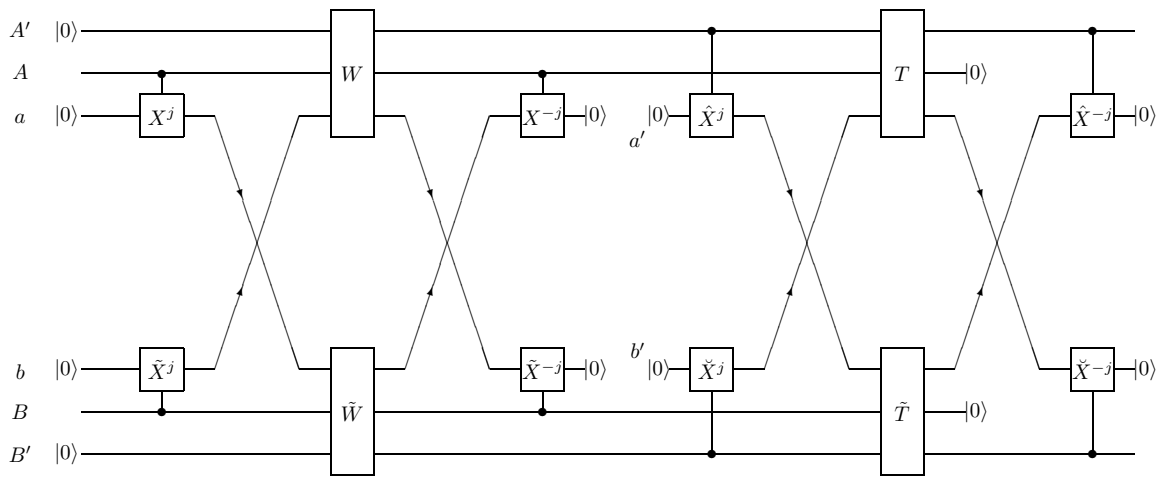}
\end{center}
\caption{The circuit diagram for Protocol~\ref{ptl3}. It implements any bipartite permutation unitary $U$ of Schmidt rank $r$ with input in $AB$ and output in $A'B'$, using LOCC and at most $8r-8$ ebits of entanglement, where $r$ is any positive integer. A solid inclined line with arrows represents teleportation. The $W,\tilde W,T,\tilde T$ are controlled permutation gates defined in the protocol. The $W$ is controlled by the systems $A$ and (teleported) $b$, and the $T$ is controlled by the systems $A'$ and (teleported) $b'$. The $a$ stores the input type of $A$ in the loose sense, while $a'$ is for the output type of $A$ in the loose sense. Hence the dimensions of $a$ and $a'$ may be unequal. Similar statements can be said for the $B$ side.} \label{fgr2}
\end{figure*}
The circuit diagram for the protocol is shown in Fig.~\ref{fgr2}. The steps of the protocol are as follows.

1. Alice prepares an ancilla $a$ in the state $\ket{0}$, and performs a controlled-$X^j$ gate on $A$ and $a$ (with projectors on $\cH_A$ of rank possibly greater than one) so that the system $a$ stores in its $Z$ basis the information about the type of input state of $A$ in the loose sense, which is defined before Lemma~\ref{le:input_types_A}. She teleports $a$ to Bob's side using prior shared entanglement and LOCC. Similarly, Bob prepares an ancilla $b$ storing the information about the type of input state of $B$ in the loose sense, and he teleports $b$ to Alice's side.

2. Alice performs a controlled-permutation unitary $W$ on $A'$, $A$ and the teleported $b$, with $A$ and $b$ as the control, and the $A'$ was initialized in $\ket{0}$ before such gate.  The controlled operator acting on $A'$ in the gate $W$ is a permutation unitary that only swaps the $\ket{0}$ state with the output state determined by the state on the control registers, and keeps other $Z$ basis states of $A'$ unchanged (those states are not the actual input state anyway). After the $W$, the $Z$-information about the output of $A$ under the action of $U$ is stored in the $Z$ basis of $A'$. Similarly, Bob performs $\tilde W$ and the $B'$ now contains the $Z$ information about the output of $B$ under $U$.

3. Alice teleports $b$ back to Bob's side, and Bob teleports $a$ back to Alice's side. Each party performs the inverse of the controlled gate in step 1 to erase the $a$ and $b$ to $\ket{0}$.

4. This step is similar to step 1, except that $U^\dag$ instead of $U$ is considered here, and the $A'$ and $B'$ are regarded as the input for the unitary $U^\dag$. An ancillary system $a'$ is initialized in $\ket{0}$, and after the controlled gate on $A'$ and $a'$, the $a'$ contains the type of state of $A'$ in the loose sense, and is teleported to the other side. Similarly, the $b'$ containing the type of state of $B'$ in the loose sense is teleported to Alice's side.

5. This step is similar to step 2. The controlled permutation gates $T$ and $\tilde T$ are defined similar to the $W$ and $\tilde W$ in step 2, but with $U^\dag$ instead of $U$ and the $A'$ and $B'$ taking the role as the input for $U^\dag$. Because of the form of the $T$ and $\tilde T$ gates and the states of $A$ and $B$ just prior to this step, the $A$ and $B$ are erased to $\ket{0}$.

6. This step is similar to step 3. Alice teleports $b'$ back to Bob's side, and Bob teleports $a'$ back to Alice's side. Each party performs the inverse of the controlled gate in step 4 to erase the $a'$ and $b$ to $\ket{0}$. This completes the protocol, with the output of $U$ in systems $A'$ and $B'$.
}
\eptl
In the protocol above, we need to erase the $A$, $B$ (which become ancillary systems in the end) and other ancillas to some fixed state, because no information about the input should be leaked to ancillas in the end; otherwise the protocol does not implement a unitary operator (c.f. \cite{ygc10}, Sec. II C). The above protocol computes the correct output states on $A'B'$ for the input computational states on $AB$ without introducing extra phases, and by linearity, it implements the unitary $U$ on all input quantum states.

\bt\label{thm:permutation}
Any bipartite permutation unitary of Schmidt rank $r$ can be implemented using local operations with the help of $\min\{\log_2(B_{r+1})+r+\log_2 r, 8r-8\}$ ebits of entanglement and twice as many c-bits.
\et

\smallskip
The proof of this theorem is given in Appendix \ref{app:{thm:permutation}}.
This significantly improves over the result in Theorem 22 of \cite{cy15}, which states that such unitary can be implemented using LOCC with $3\times 2^r$ ebits. Since $B_r<[0.792 r/\log_e (r+1)]^r$ for any integer $r\ge 1$ \cite{bt10}, the first term in the result of Theorem~\ref{thm:permutation} scales as $O(r\log(r))$, while the second term $8r-8$ scales as $O(r)$, but the first term is smaller for many integer values of $r$, at least including all $r<1100$ (note that for very small $r$, the exact value of $B_{r+1}$ is used in the calculation rather than the asymptotic bound above). Also, note that the duration of time of classical communication in Protocol~\ref{ptl2} (not including the time for entanglement preparation) could be as low as $3L/c$ (since the two teleportations from the $B$ side to the $A$ side can be done simultaneously with the sending of the classical message $m$; there are also two classical messages $l$ and $n$ sent from $A$ to $B$ before and after such step), where $L$ is the distance between the two parties, and $c$ is the speed of light. The communication time required by Protocol~\ref{ptl3} is also $3L/c$, since the middle two among the four stages of teleportations can be combined into one. Combining the considerations of entanglement cost and communication time, Protocol~\ref{ptl2} has a definite advantage over Protocol~\ref{ptl3} for small $r$. In the case $r=4$, an improved bound is provided by the following corollary:

\bcr\label{cr:sr4perm}
Any bipartite permutation unitary of Schmidt rank four can be implemented using LOCC with the help of not more than $10.71$ ebits of entanglement.
\ecr
\bpf
Denote the bipartite permutation unitary as $U$. If there is a big column of $U$ containing four nonzero blocks, from Lemma~\ref{le:persch} (iv), $U$ is a controlled permutation unitary with four terms, hence the entanglement cost is at most $\log_2 4=2$ ebits. If there is a big column of $U$ containing three nonzero blocks, from Lemma~\ref{le:persch} (vi) and (vii), the entanglement needed is not more than $\max\{2+\log_2 2, 1+\log_2 3\}=3$ ebits, under a protocol that may have up to three levels of control depending on $U$.  For the remaining cases, there is a formula $\log_2 (B_{r+1} \cdot r \cdot 2^r)$ in the proof of Theorem~\ref{thm:permutation}, and the $r$ term is now replaced with $2$ because any big column of $U$ contains at most $2$ nonzero blocks. This gives $\log_2 (52\times 2\times 16)<10.71$ ebits. Taking into consideration all cases, the entanglement cost of $U$ is not greater than $10.71$ ebits.
\epf

In Theorem~\ref{thm:classical_bipartite}(i) below, the two methods for implementing a bipartite permutation unitary in the proof of Theorem~\ref{thm:permutation} are adapted to the classical bipartite reversible circuits after simple changes. The implementation in Theorem~\ref{thm:classical_bipartite}(ii) below has some ancillas with final values not equal to initial values. Generally, in a classical computation on one party that uses reversible gates only, if it is required to restore the ancillas to their initial value in the end, we may copy the computation result by CNOT gates (the CNOT is a reversible gate) to some blank register, and the other ancillas can be restored to their initial value by running the inverse of the original reversible circuit. Such process is discussed in \cite{Bennett73}, and a significantly modified method is used in Protocol~\ref{ptl3} (modification is needed because the initial inputs are still present after the first part of the protocol, and they should be gotten rid of in the end for implementing a quantum unitary operation), helping us obtain the $8r-8$ term in the result about entanglement cost in Theorem~\ref{thm:permutation}. The Theorem~\ref{thm:classical_bipartite} (i) below can be directly adapted for quantum circuits that do not use entanglement but use nonlocal CNOT gates, as stated in (iii). In the following, a \emph{bipartite classical reversible map} is a reversible map from $n+m$ bits to $n+m$ bits, where the $n$ bits are on party $A$, and the $m$ bits are on party $B$. The matrix of such map is a permutation matrix. The \emph{Schmidt rank of a bipartite classical reversible map} is defined as the Schmidt rank of the corresponding quantum map, which is a bipartite permutation unitary and has the same matrix as the bipartite classical reversible map.

\bt\label{thm:classical_bipartite}
(i) Any bipartite classical reversible map of Schmidt rank $r$ can be implemented using classical local reversible gates and $\min\{2\lceil\log_2(B_{r+1})\rceil+2r+2\lceil\log_2 r\rceil, 8r-8\}$ classical nonlocal CNOT gates, if ancillas start with some known value and are required to be restored to the same value at the end.\\
(ii) Any bipartite classical reversible map of Schmidt rank $r$ can be implemented using classical local reversible gates and $2r-2$ classical nonlocal CNOT gates, with ancillas starting with some known value but without any requirement about their final value.\\
(iii) The assertion (i) also holds for quantum circuits, when the terms ``classical reversible map'', ``classical local reversible gates'', ``classical nonlocal CNOT'' and ``value'' are replaced by ``permutation unitary'', ``local permutation unitaries'', ``nonlocal CNOT'' and ``computational basis state'', respectively.
\et

The proof of this theorem is given in Appendix \ref{app:{thm:classical_bipartite}}.
Note that Theorem~\ref{thm:classical_bipartite}(ii) does not have a corresponding statement for the quantum permutation unitaries, because to implement a unitary operator, the ancillas at the end of the protocol should not contain information about the input, as mentioned in the proof of Theorem~\ref{thm:permutation}.
Also note that we do not know whether Theorem~\ref{thm:classical_bipartite} holds if all ancillas are required to start in some unknown state.
This kind of consideration also appears in \cite{xu2015}, which
uses the term ``borrowed bit'' to describe an ancillary bit whose initial value is not known and is returned to the initial value at the end of the computation.
On the other hand, in Theorem~\ref{thm:permutation} there is no specific requirement on the ancillas, so ancillas initialized in fixed quantum states are allowed and are actually used in the protocols in the proof.

\subsection{Examples}\label{subsec:permutation_example}

The simplest examples of Schmidt-rank-four permutation gates are the two-qubit SWAP and DCNOT (double-CNOT \cite{Collins01}) gates. In the following we show a more  nontrivial example of Schmidt-rank-four permutation gate, that can be implemented using Protocol~\ref{ptl2}. The Example~\ref{ex4} below is about a unitary which is the product of a few transpositions on the input system, where a transposition is a swap of two states among the computational basis states. Such gates are of interest for quantum computation: In quantum algorithms involving queries such as the Grover's algorithm \cite{Grover96,Grover97}, the oracle often acts nontrivially on only one or a few computational basis states, and is either a complex permutation gate or permutation gate, and in the former case it can often be implemented by a permutation gate with the help of ancilla qubit(s), which is illustrated in \cite{Grover01} in case of Grover's algorithm. We consider the problem of minimizing the entanglement cost across some bipartite cut of the whole input system. This is not only useful when the two parties are located in separated locations, but is also useful for a local quantum computer where some gates between certain sets of qubits may be harder to implement than other gates due to the design of the layout of the qubits, etc. In the latter case the CNOT-gate cost may be a more relevant measure than entanglement cost, but our protocols can easily be modified to use CNOT gates across a bipartite division of the whole system instead of using entanglement and classical communication (both cases are with the help of local gates), usually with linear overhead. An example for such overhead is in the proof of Theorem~\ref{thm:classical_bipartite}, which is for classical reversible circuits but can be immediately translated into a result for quantum circuit involving permutation gates.

\bex\label{ex4}
\rm
Suppose $U$ is a Schmidt-rank-four permutation unitary on a $5\times 6$ dimensional system. The matrix form of $U$ expressed using blocks is
\bea\label{eq:transposition3}
\left(
  \begin{array}{ccccc}
                     T_1 & T_3 & 0 & 0 & 0\\
                     T_2 & 0 & T_3 & 0 & 0\\
                     0 & T_2 & 0 & T_3 & 0\\
                     0 & 0 & T_2 & 0 & T_3\\
                     0 & 0 & 0 & T_2 & T_4\\
  \end{array}
  \right)
\eea
where $T_1=\diag(1,1,1,0,0,0)$, and
\bea\label{eq:t2t3}
T_2=\left(
  \begin{array}{cccccc}
                     0 & 0 & 0 & 1 & 0 & 0 \\
                     0 & 0 & 0 & 0 & 1 & 0 \\
                     0 & 0 & 0 & 0 & 0 & 1 \\
                     0 & 0 & 0 & 0 & 0 & 0 \\
                     0 & 0 & 0 & 0 & 0 & 0 \\
                     0 & 0 & 0 & 0 & 0 & 0 \\
  \end{array}
  \right),
\eea
and $T_3$ is the transpose of $T_2$, and
\bea\label{eq:t2t3b}
T_4=\left(
  \begin{array}{cccccc}
                     0 & 0 & 0 & 0 & 0 & 0 \\
                     0 & 0 & 0 & 0 & 0 & 0 \\
                     0 & 0 & 0 & 0 & 0 & 0 \\
                     0 & 0 & 0 & 0 & 1 & 0 \\
                     0 & 0 & 0 & 1 & 0 & 0 \\
                     0 & 0 & 0 & 0 & 0 & 1 \\
  \end{array}
  \right).
\eea
The $B$ space of $U$ is spanned by $T_1,T_2,T_3,T_4$, hence $U$ is of Schmidt rank four. The $U$ is a symmetric matrix, so it is easy to express the action of $U$ as the swapping of some pairs of computational basis states. The $U$ can be implemented using Protocol~\ref{ptl2}. The effective input dimension of system $A$ is three, because the second, third and fourth big columns of $U$ all have the same two nonzero blocks $T_2$ and $T_3$ in them, so the corresponding three computational basis states of $\cH_A$ are regarded as the same type of input state of $A$. The effective output dimension of $A$ relative to any of the input computational basis state of $\cH_A$ is two, because there are only two nonzero blocks in each big column of $U$. The effective output dimension of $B$ is two, because the first three computational basis states of $\cH_B$ appear in the output of $T_1$ and $T_2$, but not in $T_3$ or $T_4$, so these three states are counted as one type of output state of $B$, and the same holds for the last three computational basis states of $\cH_B$. Hence the Protocol~\ref{ptl2} requires $2\log_2 (3\times2\times 2)<3.59$ ebits for this $U$. In contrast, implementing $U$ using two-way teleportation (see the beginning of Sec.~\ref{sec:main}) would need $2\log_2 5 > 4.64$ ebits. This shows that Protocol~\ref{ptl2}  can sometimes be more efficient than two-way teleportation.
\eex

\subsection{Entangling power of bipartite permutation unitaries of small Schmidt rank}\label{ssec:ent_power_permutation}

To know how tight our upper bounds for the entanglement cost for bipartite permutation unitaries of small Schmidt rank are, it is helpful to know the entangling power of those unitaries, since the entangling power (the quantity $K_E$ in \cite{Nielsen03}) gives a lower bound for the entanglement cost under LOCC. Formally for a bipartite unitary $U$ acting on systems $AB$, we have
\bea\label{eq:K_e}
K_E(U) = \max_{\ket{\alpha},\ket{\beta}} E(U(\ket{\alpha}\ket{\beta})).
\eea
Here $\ket{\alpha}$ and $\ket{\beta}$ are pure states on system $A R_A$ and $B R_B$ respectively, $R_A$ and $R_B$ are local ancillas, and the $E$ is the von Neumann entropy of the reduced density matrix on one of the two systems $A R_A$ and $B R_B$. From the definition of $K_E$, we have $K_E(U)\le\log_2 r$ ebits for any $U$ of Schmidt rank $r$.

\bl\label{le:sch2perm_entpower}
The entangling power and entanglement cost of any Schmidt-rank-two bipartite permutation unitary are both 1 ebit.
\el
\bpf
From Lemma~\ref{lm:sch2} (i), up to local permutation unitaries and possibly a relabelling of the $A$ and $B$ sides, we may write the Schmidt-rank-two bipartite permutation unitary as $U=P_1\ox I_B + P_2 \ox V_B$, where $P_1,P_2$ are orthogonal projectors that add up to $I_A$, and $V_B$ is a permutation unitary satisfying that $V_B\ket{1}_B=\ket{t}_B$, where $t\ge 2$ is an integer, and $\{\ket{j}_B\}$ is the computational basis of $\cH_B$. Suppose $\ket{1}_A$ and $\ket{s}_A$ are computational basis states of $\cH_A$ in the support of $P_1$ and $P_2$, respectively, where $s\ge 2$ is an integer. Then for the input product state $\frac{1}{\sqrt{2}}(\ket{0}_A+\ket{s}_A)\ox \ket{1}$, the output is $\frac{1}{\sqrt{2}}\left(\ket{1}_A\ox\ket{1}_B+\ket{s}_A\ox\ket{t}_B\right)$ which contains 1 ebit of entanglement. On the other hand, we have commented previously that the entangling power of any bipartite unitary of Schmidt rank $r$ is at most $\log_2 r$ ebits.
This shows the entangling power of any Schmidt-rank-two bipartite permutation unitary is exactly 1 ebit.

From the basic controlled-unitary protocol and Lemma~\ref{lm:sch2}(i) (or from \cite{cy13}), the entanglement cost of any Schmidt-rank-two bipartite permutation unitary is not greater than 1 ebit.  Since the entangling power of 1 ebit provides a lower bound for the entanglement cost, the entanglement cost of any Schmidt-rank-two bipartite permutation unitary is exactly 1 ebit.
This completes the proof.
\epf

It should be noted that the ``entangling power'' in Lemma~\ref{le:sch2perm_entpower} can be understood as $K_E$ or $K_{\Delta E}$ (also defined in \cite{Nielsen03}), or the amortized $K_E$ or $K_{\Delta E}$ over many copies of the unitary, since all four quantities are lower bounds for the entanglement cost which is $1$ ebit in the current case.

As a side note, we consider the entangling power of complex bipartite permutation unitaries of Schmidt rank two. Their entangling power $K_E$ can take any value in the interval $(0,1]$ (ebit). A simplest class of examples are locally equivalent to the ones in \cite{Nielsen03}: $U=\sqrt{1-p} I\ox I + i \sqrt{p} \sigma_z\ox \sigma_z$, where $p\in (0,1]$. When the definition is extended to $p=0$, $U$ is a Schmidt-rank-one unitary, with $K_E(U)=0$. By the continuity of $K_E$ (see \cite{Nielsen03}), when $p$ is near zero, the $K_E(U)$ is near zero while $U$ is a Schmidt-rank-two diagonal unitary. When $p$ is near $1/2$,  the $K_E(U)$ is near $1$.

\textbf{Entangling power of Schmidt-rank-three bipartite permutation unitaries.}

The Schmidt-rank-three bipartite unitary $U$ cannot be on a $2\times2$ system \cite{Nielsen03}. Hence the maximum of $d_A$ and $d_B$ is at least three and it is indeed reachable. An example acting on $\bC^3\ox\bC^2$ is in \eqref{eq:uketbra11}. The structure of the Schmidt-rank-three bipartite permutation unitary $U$ has been partially investigated in Lemma \ref{lm:sch3perm} (i). The following result gives a range for the entangling power of such unitaries, although we do not know whether the lower bound is optimal. The upper bound of $\log_2 3$ ebits is likely not optimal for some unitaries, see case (I.1) in the proof.

\bpp\label{le:sch3perm_entpower}
The entangling power of a Schmidt-rank-three bipartite permutation unitary is at least $\log_2 9 - 16/9 \approx 1.392$ ebits and at most $\log_2 3\approx 1.585$ ebits.
\epp

The proof of this Proposition is in Appendix \ref{app:{le:sch3perm_entpower}}.
In the proof, the only case where the entangling power may be less than $\log_2 3$ ebits is case (I.1), in which case the entangling power of $U$ is at least $\log_2 9 - 16/9$ ebits, and such $U$ can be implemented using $\log_2 3$ ebits, while in general $U$ can be implemented using $2$ ebits, according to Lemma~\ref{lm:sch3perm}(i). Hence the gap between the entangling power and the entanglement cost of a Schmidt-rank-three bipartite permutation unitary is at most $\max\{2-\log_2 3, \log_2 3-(\log_2 9 - 16/9)\}< 0.42$ ebits.

Taking clue from the results above, we present the following conjecture:
\bcj\label{cj:entpower}
(1) The entangling power of any bipartite permutation unitary of Schmidt rank three can only take one of two values: $\log_2 9 - 16/9$ or $\log_2 3$ ebits.\\
(2) The entangling power of any bipartite permutation unitary of Schmidt rank $r$ can only be one of $f(r)$ distinct values, where $f(r)$ is a finite integer-valued function of $r$.
\ecj

Numerical calculations suggest that (1) is likely to hold. In the calculations we have assumed the most general form of initial product pure state with ancillas $a$ and $b$, whose sizes are assumed to be equal to those of the corresponding input system $A$ and $B$, respectively. The sizes of $a$ and $b$ need not be larger since it suffices to consider the Schmidt decomposition on $aA$ and $bB$, respectively.

\smallskip

\section{Conclusions}\label{sec:con}

We have improved the upper bound for the entanglement cost of bipartite unitary operators of Schmidt rank three under LOCC protocols. Lemma~\ref{lm:lm2} implies a structure theorem for Schmidt-rank-3 bipartite unitaries, as stated in Theorem~\ref{thm:sch3}. We have presented a protocol attaining the improved upper bound for the entanglement cost for such unitaries. We have also studied the structure and entanglement cost of bipartite permutation unitaries of Schmidt rank up to three, and presented two protocols for implementing bipartite permutation unitaries of arbitrary Schmidt rank, and analyzed the entanglement and classical communication costs of the protocols. These results are independent of the dimensions of the spaces that the unitary acts on, and they significantly improve over the corresponding results in \cite{cy15}. The results are applied to classical circuits for implementing bipartite permutation operations, and the protocols we found are such that whether requiring the ancillas to be restored to the fixed initial state makes a difference in the required number of nonlocal CNOT gates. As for the complex permutation unitaries, our progress is mostly restricted to Schmidt rank three (apart from some results for special cases of general Schmidt rank in Lemma~\ref{le:persch}): Any Schmidt-rank-three bipartite complex permutation unitary that is not equivalent to a diagonal unitary under local permutation unitaries can be implemented with $3$ ebits and LOCC, but it remains open whether there is a constant upper bound of entanglement cost for implementing an arbitrary Schmidt-rank-three bipartite diagonal unitary.

We also have quantified the entangling power of bipartite permutation unitaries of Schmidt rank two and three, and in the Schmidt-rank-three case the results suggest that there might be a gap between the entanglement cost and the entangling power. The examples of Schmidt-rank three bipartite permutation unitaries appearing in our proofs may be in some sense the simplest examples of a gap between the entanglement cost and the entangling power, if such gap exists at all: Although there are Schmidt-rank-two unitaries that may have such gap, those are not permutation unitaries and thus may be harder to study. Also, there is some correspondence between the permutation unitaries and the classical reversible circuits. So if the gap exists, there might be some operational implications even classically.

Looking at this gap problem from the limit of large Schmidt rank, an apparent open problem is whether the results of Theorems~\ref{thm:permutation} and \ref{thm:classical_bipartite} can be improved. It is known \cite{Lovett14} that any total boolean function of rank $r$ can be computed by a deterministic classical communication protocol with $O(\sqrt{r}\cdot\log(r))$ bits of communication. The problem of implementing bipartite permutations might be a harder problem than computing a boolean function on bipartite inputs, but it would be interesting to find out more about the relation between the two problems.

\smallskip
\section*{Acknowledgments}
L.Y. thanks Kae Nemoto for helpful discussions.
L.C. was supported by the NSF of China (Grant No. 11501024), and the Fundamental Research Funds for the Central Universities (Grant Nos. 30426401 and 30458601).
L.Y. was supported by NICT-A (Japan).

\appendix

\section{The proof of Lemma~\ref{lm:lm2}}
\label{app:{lm:lm2}}

\bpf
Firstly, note that $T_2$ must have at least two distinct eigenvalues, since otherwise $U$ is of Schmidt rank $2$, violating the assumption that it is of Schmidt rank $3$.
Another observation is that the ratio $c_{j2}/\sqrt{\vert c_{j1}\vert^2 + \vert c_{j2}\vert^2}$ (and hence $c_{j2}^\ast/\sqrt{\vert c_{j1}\vert^2 + \vert c_{j2}\vert^2}$) takes at least two different values among different $j$, since otherwise $U$ is expandable using the two operators $c_{j1}I_d+c_{j2}T_2$ and $T_3$ on the second system with any particular $j$, implying that $U$ is of Schmidt rank $2$.
Let
\bea
\label{eq:t3}
T_3=D_3+E_3,
\eea
where $D_3$ is diagonal, and all diagonal elements of $E_3$ are zero. Then $E_3$ is nonzero.
Since $U$ is unitary, \eqref{eq:u14new2} implies that
\bea\label{eq:cj_tj}
(c_{j1} I_d + c_{j2} T_2 + c_{j3} T_3)
(c_{j1} I_d + c_{j2}^\ast T_2^\dag + c_{j3}^\ast T_3^\dag)=I_d,~~\notag
\\
(c_{j1} I_d + c_{j2}^\ast T_2^\dag + c_{j3}^\ast T_3^\dag)
(c_{j1} I_d + c_{j2} T_2 + c_{j3} T_3)=I_d,~~
\eea
for all $j\in \{1,\dots,K\}$.
Given that $T_3 T_3^\dag=T_3^\dag T_3=I_d$, we subtract terms with $T_3 T_3^\dag$ or $T_3^\dag T_3$ from both sides of each equation in \eqref{eq:cj_tj}.
Since any $c_{j1}$ is real, the off-diagonal part of the resulting equations gives that
\bea\label{eq:cj_tj_2}
c_{j1} c_{j3}^\ast E_3^\dag + c_{j1} c_{j3} E_3
+
c_{j2} c_{j3}^\ast T_2 E_3^\dag + c_{j2}^\ast c_{j3} E_3 T_2^\dag =0,~~\notag
\\
c_{j1} c_{j3}^\ast E_3^\dag + c_{j1} c_{j3} E_3
+
c_{j2} c_{j3}^\ast E_3^\dag T_2 + c_{j2}^\ast c_{j3} T_2^\dag E_3 =0~~~
\eea
for all $j\in \{1,\dots,K\}$.
Since $c_{j3}$ are nonzero, we may divide both sides of the first equation in \eqref{eq:cj_tj_2} by $c_{j3}$, and obtain two independent equations of variables $E_3$ and $E_3 T_2^\dag$ by letting $c_{j2}^\ast/\sqrt{\vert c_{j1}\vert^2 + \vert c_{j2}\vert^2}$ take two different values (the other two terms containing $E_3^\dag$ and $T_2 E_3^\dag$ are viewed as ``constants''). Hence $E_3$ and $E_3 T_2^\dag$ are in the space $H:=\lin\{E_3^\dag,T_2 E_3^\dag\}$.
If $E_3^\dag\propto T_2 E_3^\dag$, then $T_2$ is proportional to the identity matrix on the rows in which $E_3^\dag$ is nonzero. The remaining diagonal elements of $T_2$ are in the rows in which $E_3^\dag$ is zero. By \eqref{eq:t3} and the unitarity of $T_3$, the columns of $E_3^\dag$ that contain these diagonal entries (at the same positions in both $T_2$ and $T_3$) are also zero. Hence $T_2$ and $T_3$ are simultaneously block-diagonal under a block structure where the first block of $T_2$ is proportional to the identity matrix. It violates the assumption that $T_2$ and $T_3$ are not simultaneously diagonalizable under a unitary similarity transform. Therefore $H$ has dimension two. We discuss two cases.

Case (a). Here $E_3^\dag$ and $E_3$ are not proportional, so they form a basis of $H$. We have $T_2 E_3^\dag=g E_3 + h E_3^\dag$ with complex numbers $g,h$. Since $E_3^\dag$ and $T_2 E_3^\dag$ also form a basis of $H$, we have $g\ne0$. Then
\bea
\label{eq:t2'}
T_2' E_3^\dg = E_3
\eea
with a diagonal matrix $T_2':=(T_2-h I_d)/g$. Denote $t_j$ as the $j$-th diagonal element of $T_2'$. It follows from \eqref{eq:t3} and the unitarity of $T_3$ that the row vector and column vector of $E_3$ containing a diagonal entry of the same position have equal norm. Let $e_{jk}$ be the $(j,k)$ element of $E_3$. Let $S:=\{j: \exists k\,\,\mbox{s.t.}\,\,e_{jk}\ne 0\}$. Then it follows from \eqref{eq:t2'} that $t_j$ for those $j\in S$ all have modulus one. It follows from \eqref{eq:t2'} that $t_j e_{kj}^*=e_{jk}$ and $t_k e_{jk}^*=e_{kj}$, $\forall j,k\in\{1,\cdots,d\}$. So if $e_{jk}\ne0$, then $j\in S$ and $t_j=t_k$. Then $t_j\ne t_k$ implies $e_{jk}=0$. The last result, combined with the definition of $T_2'$ and \eqref{eq:t3}, implies that $T_2$ and $T_3$ are simultaneously block-diagonal, where the blocks are such that each diagonal block of $T_2$ is a scalar matrix. Hence $T_2$ and $T_3$ are simultaneously diagonalizable under a unitary similarity transform. It is a contradiction with the assumption in the lemma.  So case (a) has been excluded.

Case (b). Hence $E_3^\dag$ and $E_3$ are proportional. By adjusting the phase for $E_3$, while multiplying all $c_{j3}$ by a corresponding phase factor to keep $U$ unchanged, we have $E_3^\dag=E_3$. Applying this equation to the two equations in \eqref{eq:cj_tj_2}, we have
\bea\label{eq:cj1j3}
&&-(c_{j1} c_{j3}^\ast+ c_{j1} c_{j3}) E_3
\notag\\
&=&
c_{j2} c_{j3}^\ast T_2 E_3 + c_{j2}^\ast c_{j3} E_3 T_2^\dag
\notag\\
&=&
c_{j2} c_{j3}^\ast E_3 T_2 + c_{j2}^\ast c_{j3} T_2^\dag E_3
\eea
for all $j\in \{1,\dots,K\}$. Left-multiplying the last line (which is equal to the first line) by $T_2$ and right-multiplying it by $T_2^\dg$, we obtain the second line, which is also equal to the first line, thus we have $(c_{j1} c_{j3}^\ast+ c_{j1} c_{j3}) (E_3 - T_2 E_3 T_2^\dg )=0$.
Since the unitaries $T_2$ and $T_3$ are not simultaneously diagonalizable, we have \bea
\label{eq:cj1}
c_{j1} c_{j3}^\ast+ c_{j1} c_{j3}=0,~~\forall j\in \{1,\dots,K\}.~~
\eea
Hence the first line of \eqref{eq:cj1j3} is zero, thus the second line of \eqref{eq:cj1j3} is zero, and since $c_{j2}$ and $c_{j3}$ are nonzero, we have $T_2 E_3\propto E_3 T_2^\dag$.
We may adjust the phase of $T_2$ (while multiplying all $c_{j2}$ by a corresponding phase factor) so that
\bea
\label{eq:t2e3}
T_2 E_3=- E_3 T_2^\dag.
\eea
From \eqref{eq:cj1} and the fact that all $c_{j1}$ are positive, we obtain that all $c_{j3}$ are pure imaginary.
The last two statements, combined with that the second line of \eqref{eq:cj1j3} is zero, imply that all adjusted $c_{j2}$ are also pure imaginary.

In the rest of the proof we use three assumptions. First, up to a relabeling of the computational basis states of $\cH_B$, $T_2=\bigoplus_{k=1}^n T_2^{(k)}$, and $T_3=\bigoplus_{k=1}^n T_3^{(k)}$, where $T_2^{(k)}$ and $T_3^{(k)}$ both act on the subspace $\cH_{B_k}$ of $\cH_B$, and $T_2^{(k)}$  is diagonal. Second, $T_2^{(1)}$ and $T_3^{(1)}$ commute, and the order of the matrix $T_2^{(1)}$ is the largest possible under this requirement and the first assumption. Of course it may be possible that such order is zero. If the order is nonzero, there is a  unitary change of basis in the subspace $\cH_{B_1}$, such that the transformed $T_2^{(1)}$ and $T_3^{(1)}$ are diagonal, while keeping the identity matrix in this subspace [see the $I_d$ term in \eqref{eq:u14new2}] unchanged. Third, for any $k>1$,  $T_2^{(k)}$ and $T_3^{(k)}$ do not commute, and no $T_3^{(k)}$ can be block diagonal in a basis in which $T_2^{(k)}$ is diagonal. So any $T_2^{(k)}$ with $k>1$ has at least two distinct eigenvalues. It can be easily verified that the three assumptions as a whole is always valid, although it is possible that $\cH_{B_1}$ is a null space for some $U$.

In the following derivations the $k$ is always greater than $1$ unless otherwise stated. Using \eqref{eq:t3}, we have
\bea
\label{eq:t3k}
T_3^{(k)}=D_3^{(k)}+E_3^{(k)},
\eea
where $D_3^{(k)}$ is diagonal and the diagonals of $E_3^{(k)}$ are zero. Using \eqref{eq:t2e3} and \eqref{eq:t3k}, we have
\bea
\label{eq:t2ke3}
T_2^{(k)} E_3^{(k)}=- E_3^{(k)} (T_2^{(k)})^\dag.
\eea
This equation and the assumptions imply that any $T_2^{(k)}$ has exactly two distinct eigenvalues $e^{i\a_k},-e^{-i\a_k}$ with a real number $\a_k$.
There exists a permutation matrix $P_k$ such that
\bea
\label{eq:pkt2k}
P_k T_2^{(k)} P_k^\dg
&=&
e^{i\a_k} I^{(k)}_{d_k} \oplus (-e^{-i\a_k}) I_{e_k}^{(k)}
,
\\
\label{eq:pke3}
P_k E_3^{(k)}P_k^\dg
&=&
\left(
                   \begin{array}{cc}
                     0 & F_3^{(k)} \\
                     G_3^{(k)} & 0 \\
                     \end{array}
                 \right),
\eea
where $d_k$ and $e_k$ are positive integers. Since $E_3^\dag=E_3$, we have $G_3^{(k)}=(F_3^{(k)})^\dg$. Since $T_3^{(i)}$ is unitary, \eqref{eq:t3k} implies that any two row vectors of $F_3^{(k)}$ are orthogonal, and any two column vectors of $F_3^{(k)}$ are also orthogonal. Our assumptions and the unitarity of $T_3^{(k)}$ imply that there is no zero row or column in $F_3^{(k)}$. The last two sentences imply $e_k=d_k\ge 1$.
Then the unitary
\bea
\label{eq:s2k}
S_2^{(k)}:=
I_{d_k} \oplus (-I_{d_k})\in\lin\{I^{(k)},T_2^{(k)}\}
\eea
satisfies $S_2^{(k)}=(S_2^{(k)})^\dg$.
If $d_k>1$, suppose $D_3^{(k)}$ is nonzero. Let the $V$ and $D$ in Lemma~\ref{le:linearcom} correspond to $T_3^{(k)}$ and $D_3^{(k)}$, respectively.  From the form of $T_2^{(k)}$ in \eqref{eq:pkt2k}, and noting the form of the unitary similarity transform in Lemma~\ref{le:linearcom}, it can be found that Lemma~\ref{le:linearcom} contradicts with the assumption that ``no $T_3^{(k)}$ can be block diagonal in a basis in which $T_2^{(k)}$ is diagonal.''  Hence $D^{(k)}_3=0$.
Then \eqref{eq:t3k} and \eqref{eq:pke3} imply that
\bea
\label{eq:t3ke3k}
T_3^{(k)}=E_3^{(k)}
=
\left(
                   \begin{array}{cc}
                     0 & F_3^{(k)} \\
                     (F_3^{(k)})^\dg & 0 \\
                     \end{array}
                 \right),
\eea
is a unitary matrix. Then $F_3^{(k)}$ is a unitary of order $d_k$. Let the $D$ and $\tilde U$ in Lemma~\ref{le:linearcom} correspond to $c_{j1}I_{2d_k}+c_{j2}T_2^{(k)}$ and $T_3^{(k)}$, respectively, for some $j\in\{1,\dots,K\}$, where $K$ is from \eqref{eq:u14new2}. From \eqref{eq:u14new2}, there is a nontrivial linear combination of these two matrices that is a unitary, so it corresponds to $V$ in Lemma~\ref{le:linearcom}. By noting the form of $T_2^{(k)}$ in \eqref{eq:pkt2k}, and the form of the unitary similarity transform in Lemma~\ref{le:linearcom}, and the fact that a basis in which $T_2^{(k)}$ is diagonal is also a basis in which $c_{j1}I_{2d_k}+c_{j2}T_2^{(k)}$ is diagonal, and vice versa, it can be found that Lemma~\ref{le:linearcom} contradicts with the assumption that ``no $T_3^{(k)}$ can be block diagonal in a basis in which $T_2^{(k)}$ is diagonal.'' The argument above excludes the possibility of $d_k>1$. We have $d_k=1$.

The above argument implies that $T_2^{(k)}$ in \eqref{eq:pkt2k} and $T_3^{(k)}$ in \eqref{eq:t3ke3k} are both $2\times2$ unitary matrices, and $\det T_2^{(k)}=-1$. From \eqref{eq:pke3}, by doing a conjugation by a suitable diagonal $2\times2$ unitary: $T_3^{(k)}\ra Q_k T_3^{(k)} Q_k^\dg$, we may assume that the two non-diagonal entries of $T_3^{(k)}$ are equal and positive. The conjugation by the diagonal unitary $I_{T_2^{(1)}}\op (\op^n_{k=2} Q_k)$ does not change the $I_d$ and $T_2$ in \eqref{eq:u14new2} since the latter are both diagonal. This completes the proof.
\epf

\section{The proof of Theorem~\ref{thm:sch3}}
\label{app:{thm:sch3}}

\bpf
(i) The assertion follows from the following argument which uses Lemma~\ref{lm:lm2}.

The condition that $U$ is a Schmidt-rank-3 bipartite unitary controlled from the $A$ side implies $d_A\ge 3$ and $d_B\ge 2$.
We consider the following decomposition of a general Schmidt-rank-three unitary $U$ controlled from the $A$ side:
\bea\label{eq:u_sum_T_j}
U=\sum_{j=1}^{d_A} \ketbra{j}{j} \ox T_j,
\eea
where the unitaries $T_1,T_2$ and $T_3$ are linearly independent, and other $T_j\in\lin\{T_1,T_2,T_3\}$ are unitary. Using a local unitary on $\cH_B$, we assume $T_1=I_B$. We define the set $S_1:=\{T_j: T_j\in\lin\{T_1,T_2\}\}$, thus $T_1$ and $T_2$ are in $S_1$. We refer to $S_2$ as the set of $T_j$ (including $T_3$) that are in $\lin\{T_1,T_3\}$ but not in $S_1$. We also refer to $S_3$ as the set of $T_j$ that are not in $S_1\cup S_2$. Every $T_j$ in $S_3$ is of the form $T_j=\sum_{k=1}^3 h^{(j)}_k T_k$ with nonzero $h^{(j)}_2$ and $h^{(j)}_3$. The set $\{T_j\}_{j=1}^{d_A}$ is the union of the disjoint sets $S_1$, $S_2$ and $S_3$. Let the part of unitary $U$ corresponding to the set $S_{k}$ be denoted by $W_k$, $k=1,2,3$. Using these notions and \eqref{eq:u_sum_T_j}, we have that up to a relabelling of the computational-basis states on $\cH_A$,
\bea
\label{eq:u1_1}
U=W_1 \oplus_A W_2 \oplus_A W_3.
\eea
Evidently each of $W_1$ and $W_2$ has Schmidt rank at most two, and $W_3$ has Schmidt rank at most three. Consider the following two cases.

Case (a): $W_3$ has Schmidt rank not greater than two. In this case $U$ is of the first standard form in assertion (i), according to \eqref{eq:u1_1}.

Case (b): $W_3$ has Schmidt rank exactly three. We may apply suitable local unitaries on $\cH_B$ before and after $U$ so that $T_1=I_B$ and $T_2$ is diagonal, thus in the case that  $T_2$ and $T_3$ are not simultaneously diagonal, Lemma~\ref{lm:lm2} could be applied to $W'_3=(D^{(3)}_A\ox I_B) W_3$, where $D^{(3)}_A$ is a diagonal unitary on the subspace of $\cH_A$ that $W_3$ resides in, so as to let $W'_3$ satisfy the assumption in Lemma~\ref{lm:lm2} that all $c_{j1}$ are real, thus $W'_3$ is of the second standard form in assertion (i), then so is $W_3$. The case that $T_2$ and $T_3$ are simultaneously diagonalizable is excluded in the assumptions of Lemma~\ref{lm:lm2}, but this case is possible, and $W_3$ is locally equivalent to a diagonal unitary in this case, so the second standard form in assertion (i) still holds for $W_3$. Then since $T_1,T_2,T_3$ span the $B$ space of $U$ as well as the $B$ space of $W_3$, the unitary $U$ also is of the second standard form.

(ii) Since $U$ is controlled from the $A$ side, it can be implemented using the basic controlled-unitary protocol with $\log_2 d_A$ ebits and LOCC. This gives the $d_A$ term inside the $\min\{\}$ symbol in Eq.~\eqref{eq:log2}.

The two-way teleportation protocol with the $B$ system being teleported, gives the $d_B^2$ term inside the $\min\{\}$ symbol in Eq.~\eqref{eq:log2}.

If $U$ is of the first standard form in assertion (i), the $U$ is a two-level controlled unitary, where the higher level controls which of the (up to) three unitaries $W_1,\dots,W_3$ is to be implemented in the lower level. Each of the three unitaries in the lower level is a controlled unitary of Schmidt rank two, thus there is one side in which it is controlled with two terms \cite{cy13}. Thus $U$ can be implemented under Protocol~\ref{ptl1b} with at most $\log_2 3 + 1 = \log_2 6$ ebits and at most $2\log_2 3 + 4=2\log_2 12$ c-bits. Since $d_B\ge 2$, the $\log_2 6$ ebits is not greater than the entanglement cost discussed in the next paragraph, and the relation of the entanglement costs in the current paragraph and the next paragraph is that the maximum is to be taken between these two, thus the $\log_2 6$ term does not appear in Eq.~\eqref{eq:log2}.

Now consider the second standard form in assertion (i). We may use Protocol~\ref{ptl_gp} with the choice of group being the dihedral group $D_{2n}$ with odd $n$. The group is of order $2n$, using the convention in \cite{dihedral} (note that the same group is sometimes denoted as $D_n$ in the literature). From the representation theory of dihedral groups \cite{dihedral}, such group $D_{2n}$ has $(n-1)/2$ irreducible two-dimensional representations and two one-dimensional representations. There are $\lfloor d_B/2\rfloor$ $2\times 2$ blocks and possibly a $1\times 1$ block on the $B$ side of the expansion of the bipartite unitary, by viewing (as many as possible) pairs of $1\times 1$ blocks as $2\times 2$ blocks. Thus we have $n=2\lfloor d_B/2\rfloor+1$, and the order of the group is $2n=4\lfloor d_B/2\rfloor+2$. So the group-type protocol needs $\log_2 (4\lfloor d_B/2\rfloor+2)$ ebits. The asserted entanglement-cost upper bound \eqref{eq:log2} is obtained by combining the results of the cases above.

In all the cases mentioned above except the first standard form in assertion (i), the number of bits of classical communication is twice the amount of ebits contained in the resource entangled state, thus the claim of classical communication cost in assertion (ii) holds.
\epf

\section{The proof of Lemma~\ref{le:persch}}
\label{app:{le:persch}}

\bpf
(i) Since $U$ is a complex permutation matrix, any two nonzero entries in $U$ are in different rows of $U$. So the first assertion holds. The number of the nonzero blocks cannot exceed the Schmidt rank of $U$, which is $r$. On the other hand the number cannot be zero because $U$ is unitary. So the second assertion holds.

(ii) Up to local permutation matrices, we may assume that the first $r$ blocks in the big row are nonzero, and the remaining blocks in the big row are zero.
Since $U$ is a complex permutation matrix and each block is of size $d_B\times d_B$, there are exactly $d_B$ nonzero entries in distinct rows of the big row. If the $r$ nonzero blocks in the given big row contains a common zero column vector, then any linear combination of them contains a zero column vector of the same position. And since $U$ is of Schmidt rank $r$, any block of $U$ is zero in that particular column. It is a contradiction with the fact that $U$ is unitary. So these $r$ blocks do not contain any common zero column vector. Since there are exactly $d_B$ nonzero entries in the $r$ blocks, the nonzero entries in the $r$ blocks are in different columns. So the assertion follows. Similarly, the assertion holds when all ``row'' are replaced with ``column''.

(iii) The first two sentences in the claim follow from the fact that any block in $U$ is the linear combination of the $r$ nonzero blocks described in (ii). The last sentence in the claim is reached by using the basic controlled-unitary protocol.

(iv) The argument is exactly similar to the proof of (ii)(iii), so we abbreviate it here.

(v) Up to local permutation unitaries, we may assume that the first big row of $U$ contains exactly $r-1$ nonzero blocks, and the nonzero blocks in it are the first $r-1$ blocks, with the first one being equal to $I_s \oplus 0_{d_B-s}$, where $1\le s\le d_B-r+2$. In the following we prove that up to local permutations all the $r-1$ nonzero blocks in the first big row can be written as orthogonal projectors. Suppose this were not true, then there would be at least one common zero column in these $r-1$ nonzero blocks, and the $r$-th linearly independent block in $U$ must contain a nonzero element in this column. The linear combination of the $r$-th block and the first $r-1$ blocks (with nonzero coefficient for the $r$-th block) can appear at most once in each big row except the first big row, but must appear in each big column. Thus the count of such linear combination is both not more than $d_A-1$ and exactly equal to $d_A$, and this is a contradiction. Hence, up to local permutations all the $r-1$ nonzero blocks in the first big row can be written as orthogonal projectors. So the assertion holds.

(vi) The conditions imply that $\sum^{r-1}_{j=1} P_j = I_B$, $\sum^{r-1}_{j=1} s_j = d_B$, and the sum of orders of $Q$ and $Q_j$ ($\forall j\le n$) is $d_A$. These facts are used in the following proof.

Since $U$ has Schmidt rank $r$, there is the $r$'th linearly independent block in $U$. This is a complex partial permutation matrix named as $R$. We regard it as a partitioned matrix
\bea
\label{eq:rsum}
R:=\sum^{r-1}_{j,k=1}\ketbra{j}{k}\ox R_{jk},
\eea
where the subblock $R_{jk}$ is of size $s_j \times s_k$. In particular, the diagonal subblock $R_{jj}$ is in the same position and of the same size as that of $P_j$ in any diagonal block of $U$.
Up to local permutation matrices on $U$, we may use the hypothesis that $n$ is the integer such that for any $j\in[1,n]$, there is a nonzero $R_{j,k_1}$ or $R_{k_2,j}$; and at the same time any $R_{j,k_1}$ and $R_{k_2,j}$ are both zero when $j>n$, $j\ne k_1,k_2$, and $k_1,k_2\in[1,r-1]$.
In other words, $R$ is the direct sum of the upper left
$(\sum^n_{j=1} s_j)\times(\sum^n_{j=1} s_j)$ submatrix $R'$ and $r-n-1$ subblocks $R_{jj}$ of size $s_j \times s_j$, $j=n+1,\cdots,r-1$, where the integer $n\in\{0\}\cup\{2,\dots,r-1\}$, since $n=1$ implies that there is a nonzero off-diagonal block $R_{1k}$ where $k\ge 2$, meaning that $n\ge 2$, thus the case $n=1$ does not exist.

Since $U$ is a complex permutation matrix, any block of $U$ is a complex partial permutation matrix, which is the linear combination of $P_1,\cdots,P_{r-1}$ and $R$. These facts, \eqref{eq:rsum} and the hypothesis imply that $P_1,\cdots,P_n$ do not appear in the linear combination containing $R$ of nonzero coefficient. So any block in $U$ is either the linear combination of $P_1,\cdots,P_{r-1}$, or the direct sum of $R'$ multiplied by a phase and $r-n-1$ subblocks of size $s_j \times s_j$, $j=n+1,\cdots,r-1$. The hypothesis implies that each subblock is the linear combination of $R_{jj}$ and $P_j$. The submatrix of $U$ on the bipartite Hilbert space $\cH_A\times \lin\{\ket{s_1+\cdots+s_n+1},\cdots,\ket{d_B}\}$ form the second bracket in \eqref{eq:ubigg}. The remaining part of unitary $U$, named as $U'$, acts on the bipartite Hilbert subspace $\cH_A\times\lin\{\ket{1},\cdots,\ket{s_1+\cdots+s_n}\}$. The above argument implies that each block of $U'$ is the linear combination of $P_1,\cdots,P_n$ and $R'$. In particular, the block has to be proportional to $R'$ when $R'$ appears in the linear combination. The hypothesis also implies that the big row or big column of $U'$ containing $R'$ does not contain any other nonzero block. So $R'$ is a complex permutation matrix.
By letting $R'=P$, we can decompose $U'$ into the expression in the first bracket of \eqref{eq:ubigg}. For the $U_j$ in the last big bracket in \eqref{eq:ubigg}, it has Schmidt rank at most two, since $R$ and the term with the specific $P_j$ (where $j>n$) each contributes at most $1$ to the Schmidt rank. So the first paragraph in the claim holds.

The last paragraph in the claim is from a multiple-level recursive control protocol generalized from Protocol~\ref{ptl1}. In the case $n\in[2,r-2]$, the protocol has three levels. The first level is choosing between the two terms in \eqref{eq:ubigg}. If the choice is the first term, the second level then chooses between the two terms in the first big bracket in \eqref{eq:ubigg}. Otherwise, the second level chooses between the terms in the last big bracket in \eqref{eq:ubigg}, and the third level implements a Schmidt-rank-two unitary using the basic controlled-unitary protocol. In the case $n=0$, the protocol similarly has three levels but the first branch in the choices does not have the second or third level. In the case $n=r-1$, the protocol has only two levels since the last term in \eqref{eq:ubigg} does not exist. In all cases, the lowest level of the protocol is the basic controlled-unitary protocol.

(vii) Let $U$ be a real permutation matrix and let it be of the form of the $n=0$ case in \eqref{eq:ubigg}. We can instead expand the $U$ using orthogonal projectors $P_1, P_2, \dots, P_{r-1}$, and the matrix $R$ on the $B$ side, where $R$ is defined in the proof of (v) and is block-diagonal in the sense that $R_{jk}=0$ for $j\ne k$ in \eqref{eq:rsum}, since $n=0$. If $R$ is a diagonal matrix, then $R$ cannot be the identity matrix since then it would be in $\lin\{P_1, \dots, P_{r-1}\}$, violating that $U$ is of Schmidt rank $r$. But $R$ can be of less than full rank under the assumption that $R$ is diagonal, and in such case $U$ is a controlled permutation matrix controlled from $B$ side with at most $2(r-1)$ terms, which is the second form for $U$ in the assertion. Now suppose $R$ is not diagonal. Any block in $U$ cannot be a linear combination of $R$ and $P_1, \dots, P_{r-1}$ with nonzero coefficient for $R$, since then it would have two nonzero elements in some row. Thus any block of $U$ must be either $R$ or a linear combination of $P_1,\dots\,P_{r-1}$. Thus $U$ is the $A$-direct sum of a unitary whose $B$ space is spanned by $R$ only, and another unitary whose $B$ space is spanned by $P_1,\dots,P_{r-1}$, and the latter is a $(r-1)$-term controlled-permutation unitary controlled from the $B$ side. This is exactly the form for the case $n=r-1$ in \eqref{eq:ubigg}. Thus the assertion holds, and the statement about entanglement cost follows from Protocol~\ref{ptl1}.

This completes the proof.
\epf

\section{The proof of Lemma~\ref{lm:diagonal_blocks}}
\label{app:{lm:diagonal_blocks}}

\bpf
(i) The claim holds by definition.\\
(ii) The equality obviously holds when $r=1$. In the following we assume $r\ge 2$.
Denote the unitary as
\bea
\label{eq:usum}
U=\sum_{j=1}^{d_A} \ketbra{j}{j} \ox V_j.
\eea
A class of examples $U$ with $2^{r-1}$ distinct diagonal blocks satisfy
$d_A=2^{r-1}$, $d_B=2r-2$, $V_1=I_{2r-2}$, and for $k=2,\dots,r$, $V_k:=I_{2r-2}+\ketbra{2k-3}{2k-2}+\ketbra{2k-2}{2k-3}-\ketbra{2k-3}{2k-3}-\ketbra{2k-2}{2k-2}$. The $2^{r-1}$ diagonal blocks $V_j$ are of the form $V_r +\sum_{k=2}^r y_k (V_k-V_r)$, where $y_k$ is $0$ or $1$ for each $k\in[2,r]$. Hence
\bea
\label{eq:2r-1}
m(r)\ge 2^{r-1}.
\eea

Now we proceed with the main proof. Up to local permutations on $\cH_A$, we may assume the first $r$ diagonal blocks of $U$ in \eqref{eq:usum} are linearly independent. We still denote them as $V_1, V_2, \dots, V_r$. Since each $V_h$ is the linear combination of them, we have $V_h=\sum_{k=1}^r x^{(h)}_k V_k$. Since all $V_j$ are permutation matrices, the sum of elements in each row of any $V_j$ is $1$. Thus we have $\sum_{k=1}^r x^{(h)}_k=1$. These two equations imply
\bea
\label{eq:vh}
V_h-V_1=\sum_{k=2}^r x^{(h)}_k (V_k-V_1).
\eea
For each $k=2,\cdots,r$ we regard the $V_k-V_1$ as a $d_B^2$-dimensional vector. Let the $d_B^2\times (r-1)$ matrix $M$ be consisted of column vectors $V_2-V_1,\cdots,V_r-V_1$. Since $V_2,\cdots,V_r$ are linearly independent, $M$ is of full rank $r-1$. Since the entry sum in each row of the matrix $V_k-V_1$ is zero, we can perform fixed row operations on $M$, to make zero the $d_B$ rows corresponding to the nonzero entries of $V_1$. The resulting matrix $M'$ has the same rank as $M$, since row operations preserve the matrix rank. There is a matrix $M''$ which is a $(r-1)\times (r-1)$ submatrix of $M'$, obtained by deleting the $d_B$ zero rows and some other rows in $M'$, which has the same rank as $M$, namely $r-1$. Then \eqref{eq:vh} is equivalent to the fact that the vector $M''\cdot [x^{(h)}_2,\cdots,x^{(h)}_r]^T$ has entries one or zero, since all entries of $V_h$ are $0$ or $1$, and the nonzero entries of $V_1$ are excluded by the deletion mentioned above. So there are at most $2^{r-1}$ sets of solutions of $x^{(h)}_2,\cdots,x^{(h)}_r$. It implies $m(r)\le 2^{r-1}$. Combining it with \eqref{eq:2r-1} we have $m(r)=2^{r-1}$.

(iii) The claim follows from (ii) and the basic controlled-unitary protocol.

(iv) A set of $B$-side Schmidt operators of $U$ can be chosen to be a set of linearly independent $d_B\times d_B$ blocks in the matrix $U$, hence they are partial permutation matrices (but in general they cannot be an arbitrary set of partial permutation matrices, since they jointly have to have support on every input computational-basis state). Then the assertion follows by definition.

(v) If $m'(r)>2^{r-1}$, by assertions (i), (ii) and that $r\ge 1$, there must be at least $r$ linearly independent ones among these $m'(r)$ distinct permutation matrices. Then assertion (ii) implies $m'(r)=2^{r-1}$, a contradiction. Hence $m'(r)\le 2^{r-1}$. But by definition $m'(r)\ge m(r)$, hence $m'(r)=2^{r-1}$.

(vi) The following argument is almost the same as the last paragraph of the proof of Lemma 21 in \cite{cy15}. For completeness we include the rewritten argument below.

Suppose $\{F_i\}_{i=1}^r$ is a set of $r$ linearly independent matrices among the blocks of $U$. All nonzero partial permutation matrices in the $B$-space of $U$ are linear combinations of $\{F_j\}_{j=1}^r$. This last property still holds if we replace $\{F_i\}_{i=1}^r$ with $\{G_i\}_{i=1}^r$, defined as follows: Each $G_i$ is a linear combination of $\{F_j\}_{j=1}^r$, and satisfies $G_i(t)=\delta_{it}$, $i,t\in\{1,2,\dots,r\}$, where $G_i(t)$ is the $t$-th matrix element of $G_i$ according to some fixed ordering of the matrix elements, and $\delta_{it}$ is the Kronecker delta. Such ordering of the matrix elements must exist but the exact choice depends on the set $\{F_i\}_{i=1}^r$. We do not have extra restrictions on the $G_i(t)$ with $t>r$. Any nonzero partial permutation matrices in the $B$-space of $U$ is a linear combination of $G_i$ ($i=1,2,\dots,r$), and the coefficient for each $G_i$ is either $0$ or $1$, since the resulting matrix is a partial permutation matrix which implies that its first $r$ elements (in the ordering above) must be either $0$ or $1$. Since we only consider the nonzero matrices, the coefficients cannot all be zero, thus there are at most $2^r-1$ nonzero partial permutation matrices in the $B$-space of $U$. This proves $n(r)\le 2^r-1$.

The value $2^r-1$ is attained by a $r$-term controlled unitary controlled from the $B$ side. To prove that no other type (up to local permutation equivalence) of bipartite permutation unitaries $U$ can achieve the value $2^r-1$, we make use of the essence of the argument in the last paragraph of the proof of Lemma 21 in \cite{cy15}, that is, there are $r$ positions in the $d_B\times d_B$ matrix such that the value of these elements (each is $0$ or $1$, and is called a ``key bit'' below) determine the values of other entries of the matrices in the $B$ space of $U$ via fixed linear relations. Since there are $2^r-1$ nonzero partial permutation matrices in the $B$ space of $U$, it must be that every binary combination of the values of the $r$ key bits except the all-zero combination appear in a partial permutation matrix in the $B$ space of $U$. (Note that if the number $2^r-1$ were a smaller number, in general any binary combination of the values of the $r$ key bits does appear in some matrix in the $B$ space of $U$ but such matrix might not be a partial permutation matrix.) Thus no two key bits are located in the same row or column, since otherwise the matrix corresponding to the two key bits being both $1$ cannot be a partial permutation matrix. Suppose one key bit is at position $(r_1,c_1)$, i.e. row $r_1$  and column $c_1$, and another key bit is at position $(r_2,c_2)$. By considering the $(i,j)$ entry of the $d_B\times d_B$ matrix corresponding to both key bits being set to $1$, where $(i,j)\ne (r_1,c_1)$ and $(i,j)\ne (r_2,c_2)$, we find that such $(i,j)$ entry cannot be both $1$ in the two matrices corresponding to the two key bits being set to $1,0$ and $0,1$, respectively, as the latter two matrices add up to the former matrix. This shows that any matrix corresponding to only one key bit set to $1$ must be orthogonal to any other such matrix, where orthogonal means having no common nonzero rows and no common nonzero columns. And since the $U$ is unitary, for any row and column in the $d_B\times d_B$ matrix there has to be at least one nonzero element appearing in a partial permutation matrix with only one key bit set to $1$, thus the bipartite permutation unitary is equivalent to a controlled unitary from the $B$ side under local permutation unitaries. This completes the proof.
\epf

\section{The proof of Lemma~\ref{lm:sch3perm}}
\label{app:{lm:sch3perm}}

\bpf
(i) We call the first statement the ``assertion.'' In the following we prove the assertion first, then prove the statement about entanglement cost at the end.

We use the same notations as in the proof of Lemma~\ref{lm:sch2}.
First, if there is a big row or column of $U$ containing three nonzero blocks, then from Lemma \ref{le:persch} (iv), $U$ is equivalent to a three-term controlled-permutation unitary controlled from the $B$ side, up to local permutation unitaries.

Next, if there is exactly one nonzero block in each big row of $U$, then up to local permutation unitaries, $U$ is equivalent to a controlled-permutation unitary controlled from the $A$ side. The number of terms is between the Schmidt rank $r$ and $2^{r-1}$ by Lemma~\ref{lm:diagonal_blocks} (ii). So it is either three or four.

The remaining case is that there is a big row of $U$ containing exactly two nonzero blocks.
From Lemma~\ref{le:persch} (vi), we have a standard form in \eqref{eq:ubigg}, which satisfies the assertion except in the case $n=0$. In the case $n=0$, the assertion follows from Lemma~\ref{le:persch} (vii).

Now we prove that the entanglement cost is at most $2$ ebits. In the first case in the assertion, the result follows from the basic controlled-unitary protocol. In the only remaining case in the assertion, the result follows from Protocol~\ref{ptl1}, where the higher level of this two-level protocol determines which of the product permutation unitary or the two-term controlled-permutation unitary is to be implemented in the lower level. The entanglement cost for the two-level protocol is $\log_2 2+\log_2 2=2$ ebits. For each ebit used in the protocols, two c-bits are used, hence the classical communication cost is not more than $4$ c-bits. So the assertion holds.

(ii) Suppose $U$ is a Schmidt-rank-three bipartite complex permutation unitary that is not equivalent to a diagonal unitary under local permutation unitaries.
It follows from Lemma \ref{le:persch} (i) that some big column or row of $U$ contains the number of at most three nonzero blocks. If the number is exactly three or two, then the assertion respectively follows from Lemma \ref{le:persch} (iii) or (vi). It remains to investigate the case when the number is one.
We exchange the $A$ and $B$ systems of $U$ to obtain another matrix $\tilde U$, which is still a Schmidt-rank-three bipartite complex permutation unitary.
Since $U$ is not equivalent to a diagonal unitary under local permutation unitaries, the nonzero blocks of $U$ do not have the same nonzero patterns (the pattern about which of the elements are nonzero), hence there are two nonzero blocks of $U$ such that there is nonzero element located in the same row within each block but at different column positions. This means that some big row of $\tilde U$ contains at least two nonzero blocks. The assertion again follows from Lemma \ref{le:persch} (iii) and (vi).

(iii) Let the unitary be $U=\proj{1}\ox I_B + \proj{2}\ox (x P + y P^\perp) + \sum^{d_A}_{j=3}\proj{j}\ox V_j$ with two different phases $x,y$ and a projector $P$ onto some states in the computational basis of $\cH_B$, which can be assumed to be the first states in the basis, i.e. their labels are before the states in the support of the projector $P^\perp:=I_B-P$. The $V_j$ are diagonal matrices. We have $U=U_1 \oplus_B U_2$, where
\bea
\label{eq:u1u2}
U_1 &=& (\proj{1}+x\proj{2})\ox P + \notag\\
&&\sum^{d_A}_{j=3} \proj{j}\ox P V_j P,\notag\\
U_2 &=& (\proj{1}+y\proj{2})\ox P^\perp + \notag\\
&&\sum^{d_A}_{j=3} \proj{j}\ox P^\perp V_j P^\perp,
\eea
Since $U$ is of Schmidt rank $3$, there is a $V_j$ (denoted $V_3$ without loss of generality) that is not a linear combination of $I_B$ and $x P + y P^\perp$. Every other $V_j$ is in $\lin\{I_B,x P + y P^\perp,V_3\}$. The matrices $P V_3 P$ and $P^\perp V_3 P^\perp$ are diagonal.

If $P V_3 P$ has three or more distinct nonzero diagonal elements, then among the matrices $P V_j P$ there cannot be any linear combination of $P$ and $P V_3 P$ with nonzero coefficients for both terms, because of Lemma~\ref{le:vandermonde}(iii) and the fact that the set $\{P V_j P\}\cup \{P\}$ contains exactly two linearly independent matrices, the latter is because the set $\{V_j\}\cup \{I_B,x P + y P^\perp\}$ which span the $B$ space of the Schmidt-rank-three unitary $U$ contains exactly three linearly independent matrices. Thus every $V_j$ is either proportional to $V_3$, or is in $\lin\{I_B,x P + y P^\perp\}$. Thus $U$ can be written as $U=W_1\oplus_A W_2$, where $W_1$ is a Schmidt-rank-two unitary with the $B$ space being $\lin\{I_B,x P + y P^\perp\}$, and the $W_2$ is a product unitary with the $B$ space being spanned by $V_3$. Thus $U$ can be implemented using Protocol~\ref{ptl1}, with the lower level of this two-level protocol using at most $1$ ebit of entanglement, and the higher level (choosing between $W_1$ and $W_2$) using $1$ ebit. Thus $U$ can be implemented by $2$ ebits and LOCC in this case.

If $P^\perp V_3 P^\perp$ has three or more distinct nonzero diagonal elements, we similarly have that $U$ can be implemented with $2$ ebits and LOCC.

Now suppose $P V_3 P$ and $P^\perp V_3 P^\perp$ each has at most two distinct nonzero diagonal elements. Apparently any $P V_j P$ is in $\lin\{P,P V_3 P\}$, thus the $U_1$ (not $U$) is a unitary of Schmidt rank one or two, and can be written in a form of being controlled from the $B$ side (up to local unitaries) with at most two controlling terms. Similarly, by considering $P^\perp V_3 P^\perp$, we get that $U_2$ is controlled from the $B$ side (up to local unitaries) with at most two controlling terms. And since $U=U_1 \oplus_B U_2$, the $U$ is locally equivalent to a controlled unitary with at most $4$ controlling terms on the $B$ side, hence $U$ can be implemented using at most $2$ ebits and LOCC under the basic controlled-unitary protocol.

Hence, in all cases, the $U$ can be implemented by $2$ ebits and LOCC.
\epf

\section{The proof of Theorem~\ref{thm:permutation}}
\label{app:{thm:permutation}}\

\bpf
Denote the unitary as $U$. We first prove for the term $\log_2(B_{r+1})+r+\log_2 r$ in the assertion. In the following we consider the cases that $r\ge 4$, and the method is just to apply Protocol~\ref{ptl2} to the unitary $U$. The cases of $r\le 3$ will be mentioned later.

The dimension of $a$ (and $e$, $e'$) in Protocol~\ref{ptl2} is the effective input dimension of $A$, i.e., number of different input types of $A$, or the number of different big columns of $A$ characterized by the set of nonzero blocks in the big column regardless of the order of the blocks. The effective input dimension of $A$ is at most $B_{r+1}$, which follows from Lemma~\ref{le:combinations_partial_permutation} by noting the following: All the blocks of $U$ are in the linear span of $r$ linearly independent blocks in $U$, and we may regard the $S$ in Lemma~\ref{le:combinations_partial_permutation} as the set of all blocks in $U$, and each big column of $U$ corresponds to a covering subset of $S$ determined by which nonzero blocks are in the big column.

The dimension of $f'$ in Protocol~\ref{ptl2} is the effective output dimension of $A$ relative to the input computational basis state of $\cH_A$, and it is at most $r$, because there can be at most $r$ nonzero blocks in a big column of $U$.

The dimension of $h'$ in Protocol~\ref{ptl2} is the effective output dimension of $B$. In Def.~\ref{def_state_types}(iii) it is shown that the simplified definition is equivalent to the original definition for the effective output dimension of $B$, thus there are at most $2^r$ output types of $B$. It may be worth noting that the definition of such output types of $B$ above is independent of the output of $A$, and this is for the final phase correction $\hat Z^{-n}$ in Fig.~\ref{fgr1} to be successfully carried out.

Thus, when $r\ge 4$, the number of ebits needed in the whole protocol is at most $\log_2 (B_{r+1} \cdot r \cdot 2^r)=\log_2 B_{r+1} + r + \log_2 r <\log_2 [0.792 r/\log_e (r+1)]^r + r + \log_2 r=O(r \log r)$. For each ebit in the protocol, $2$ c-bits are needed.

When $r\le 3$, the number of ebits needed are $0$, $1$, and $2$ ebits for $r=1,2,3$, respectively, where the latter two results are from Lemma~\ref{lm:sch2} and \ref{lm:sch3perm}, respectively. Again, for each ebit in the protocols, $2$ c-bits are needed.

The above shows that $U$ can be implemented using at most $\log_2(B_{r+1})+r+\log_2 r$ ebits and twice as many c-bits.

In the following we prove for the term $8r-8$ in the assertion. From Lemma~\ref{le:input_types_A} and the symmetry of the two sides, the number of possible input types in the loose sense on each of the $A$ and $B$ sides is not more than $2^{r-1}$. Consider the Protocol~\ref{ptl3} shown in Fig.~\ref{fgr2}. The $a$ contains the input type of system $A$ in the loose sense, so its dimension is at most $2^{r-1}$. Hence the teleportation of $a$ to Bob's side requires at most $r-1$ ebits and $2r-2$ c-bits. Similarly, the teleportation of $b$ to Alice's side requires at most $r-1$ ebits and $2r-2$ c-bits. Teleporting these systems back requires the same amount of nonlocal resources.
Since $U^\dag$ has the same Schmidt rank as $U$, the entanglement and classical communication cost of the second part of the protocol is bounded above by the same numbers as in the first part of the protocol. Hence, $8r-8$ ebits and $16r-16$ c-bits suffice to implement the $U$.

Thus the assertion is proved by combining the upper bounds for the two protocols above.
\epf

\section{The proof of Theorem~\ref{thm:classical_bipartite}}
\label{app:{thm:classical_bipartite}}

\bpf
(i) For $r=2$ and $r=3$, we use the basic controlled-unitary protocol or the recursive controlled protocol [Protocol~\ref{ptl1}(a)] which are used in the proof of Lemma~\ref{lm:sch2}(i) and \ref{lm:sch3perm}(i), respectively, but with modifications to use nonlocal CNOT gates instead of entanglement, similar to those below for the case of general $r$. For $r\ge 4$, we use the adapted versions of the two protocols in the proof of Theorem~\ref{thm:permutation}. The details would be given in the following paragraphs but the main idea is to use local classical reversible gates instead of the local quantum permutation gates, and replace the entangled state and teleportation and the directly related LOCC operations with the classical nonlocal CNOT gate. According to the definition of ebits in Sec.~\ref{sec:pre}, the non-integer entanglement cost in Theorem~\ref{thm:permutation} means that a maximally entangled state on $k\times k$ system is used, where $k$ is not a power of $2$. Since we are concerned with the CNOT gate cost, we extend such entangled state to be a maximally entangled state on a $2^n\times 2^n$ system, where $n\in\mathbb{N}$, and this gives the ceiling function in the assertion. In the following we consider the two protocols in the proof of Theorem~\ref{thm:permutation} respectively.

For the first protocol in the proof of Theorem~\ref{thm:permutation} which is Protocol~\ref{ptl2}, we may use an integer number of nonlocal CNOT gates to prepare $e'$ on the $B$ side, where $e'$ is the input to the $W$ gate in Protocol~\ref{ptl2}, and similarly the same number of nonlocal CNOT gates is needed later to erase the $e'$, so two nonlocal CNOT gates are needed for every ebit in the $ee'$ state in Protocol~\ref{ptl2}. The teleportation of qubits from the $B$ side to the $A$ side are replaced with an integer number of the classical DCNOT (double-CNOT, see the quantum version in \cite{Collins01}) gates, where each DCNOT gate includes a CNOT gate controlled from $B$, where the controlled bit on $A$ is an auxiliary bit initially in the fixed value $0$, followed by a CNOT gate controlled from $A$. In other words, two nonlocal CNOT gates are used to transfer each bit from $B$ to $A$ while sending the auxiliary bit initialized in $0$ from $A$ to $B$. The original teleportation needs one ebit to teleport each qubit. Thus each term in the expression for the number of required nonlocal CNOT gates is at most two times the ceiling function of the number of ebits used in the corresponding part of Protocol~\ref{ptl2}.

For the second protocol in the proof of Theorem~\ref{thm:permutation} which is Protocol~\ref{ptl3}, each ebit can be turned into one nonlocal CNOT gate. For example, the first teleportation of the $a$ ($b$) system can be implemented by at most $r-1$ nonlocal CNOT gates to send the information about the computational basis of the register $a$ ($b$), and the teleportation back later can be implemented by at most $r-1$ CNOT gates to erase the state on one party, and then the remaining local copy of $a$ ($b$) can be locally erased by the inverse circuit of the local circuit used to prepare it. Thus the number of nonlocal CNOT gates needed is equal to the number of ebits used in Protocol~\ref{ptl3}. This completes the proof of (i).
\\
(ii) The following is the classical version of the first part of Protocol~\ref{ptl3}. From Lemma~\ref{le:input_types_A} and the symmetry of the two sides, the number of possible input types in the loose sense on each of the $A$ and $B$ sides is not more than $2^{r-1}$. Consider a classical circuit where Alice sends the input type of system $A$ to the $B$ side using $r-1$ CNOT gates, and Bob sends the input type of $B$ to the $A$ side using $r-1$ CNOT gates. Then each party computes the output of the local system, while keeping a copy of the inputs (both the local input and the received information about input types on the other system), in order to make the local circuit reversible, but this leaves some local ancillas with some value dependent on the inputs. Hence $2r-2$ CNOT gates suffice under the condition in the assertion.\\
(iii) The assertion follows from (i) as well as the fact that in the circuits in the proof of (i), the ancillas in the end do not contain information about the input. This last condition about the final state of ancillas is necessary for implementing a quantum unitary operation, and is actually sufficient as long as there are no measurements and all gates are unitary; see Theorem~1 of \cite{ygc10}.
\epf

\section{The proof of Proposition~\ref{le:sch3perm_entpower}}
\label{app:{le:sch3perm_entpower}}

\bpf
The upper bound follows from the definitions of the entangling power and the Schmidt rank of the bipartite unitary.
To prove the lower bound, we consider three possible forms of $U$, which are studied in detail below. In all cases except case (I.1), the entangling power is $\log_2 3$ ebits.

Case (I). Suppose $U$ is a controlled permutation unitary with three terms, and is controlled from the $A$ side. Up to local permutation matrices, we may assume
\bea
\label{eq:ud1}
U
&=& D_1\ox I_B
\notag\\
&+& D_2 \ox (I_m \op I_n \op V_1 \op V_2)
\notag\\
&+& D_3 \ox (I_m \op V_3 \op I_q \op V_4),
\eea
where $D_jD_k=\d_{jk} D_j$, $\sum_j D_j = I_A$, and $V_1,V_2,V_3$ and $V_4$ are permutation matrices. $V_1$ and $V_3$ are respectively of size $q\times q$ and $n\times n$, and $V_2$ and $V_4$ are both of size $p\times p$ where $p=d_B-m-n-q$. If $V_1$ or $V_3$ contains a nonzero diagonal entry, then we can move the entry by local permutation matrices on $\cH_B$ so that $I_m$ is replaced with $I_{m+1}$. So $V_1$ and $V_3$ do not contain any nonzero diagonal entry. Similarly, we may assume that $V_2$ and $V_4$ do not have a nonzero diagonal entry in the same column when $p>0$. For the purpose of studying the entangling power of $U$, we may assume that all $D_j$ in \eqref{eq:ud1} are one-dimensional projectors, since the input state is a product state.

In the following we consider three cases. The first case (I.1) is that $p=0$, namely $V_2$ does not exist in \eqref{eq:ud1}.
We perform $U$ on the product vector $\ket{e}(\ket{a}+\ket{b}+\ket{c})$ where $\ket{a}$, $\ket{b}$, and $\ket{c}$ are respectively in the support of $I_m$, $I_n$ and $I_q$ in \eqref{eq:ud1}. If the resulting state is maximally entangled, then the three states $\ket{a}+\ket{b}+\ket{c}$, $\ket{a}+\ket{b}+V_1\ket{c}$, and $\ket{a}+V_3\ket{b}+\ket{c}$ are pairwise orthogonal. The solution is $\ket{a}=\ket{b}=\ket{c}=0$.
It is a contradiction with the resulting maximally entangled state. Hence, if ancillas are not allowed, then $U$ cannot create $\log_2 3$ ebits. The unitary $U$ with $m=p=0$ and $n=q=2$ can generate $\log_2 9 - 16/9$ ebits of entanglement starting from a product state without ancillas. A corresponding choice of such input state is $\frac{1}{\sqrt{3}}(1,1,1)\otimes (g,h,g,h)$, where $g=\frac{\sqrt{3}+\sqrt{6}}{6}$, $h=\frac{\sqrt{3}-\sqrt{6}}{6}$. Numerical evidence suggests that this number of $\log_2 9 - 16/9 \approx 1.392$ ebits is optimal for this $U$, even when ancillas are allowed. Of course, if $m>0$ in the case above, we can still create the same amount of entanglement by letting the input state have zero amplitude in the support of the $I_m$. When $q$ or $n$ is greater than $2$, up to local permutations there is always an $s\times s$ cyclic shift submatrix $V_{11}$ in $V_1$ and a $t\times t$ cyclic shift submatrix $V_{31}$ in $V_3$, respectively. We ignore the $I_m$ and other parts of $V_1$ and $V_3$, which means the $B$-side input state has zero amplitude in the support of those matrices. Under these conventions, we choose the input state to be of the form $\frac{1}{\sqrt{3}}(1,1,1)\otimes (v_1,v_2)$, where $v_1$ and $v_2$ are vectors of length $t$ and $s$, respectively, and the elements in $v_1$ are just two real numbers appearing alternately: $e,f,e,f,\dots$, and thus the last number in $v_1$ is $e$ if $t$ is odd, and is $f$ if $t$ is even. Similarly the elements in $v_2$ are just two real numbers appearing alternately: $g,h,g,h,\dots$, and thus the last number in $v_2$ is $g$ if $s$ is odd, and is $h$ if $s$ is even. With suitable choices of real numbers $e,f,g,h$, this would give rise to $\log_2 9 - 16/9 \approx 1.392$ ebits of entanglement in the output state. A class of choices of the real $4$-tuple $(e,f,g,h)$ for arbitrary $t,s\ge 2$ is given by $e-f=\frac{2}{\sqrt{6\lfloor t/2\rfloor}}$, $g-h=\frac{2}{\sqrt{6\lfloor s/2\rfloor}}$ and $\vert v_1\vert=\vert v_2\vert=\frac{1}{\sqrt{2}}$. When these equations are satisfied, the output reduced density operator on the $A$ side would be determined, and is the same as that corresponding to the optimal output entangled state in the case $t=s=2$. It is not hard to see that there are two solutions for the pair $(e,f)$ and two solutions for the pair $(g,h)$ for the equations above, thus there are four solutions $(e,f,g,h)$ for these equations, for any $t$ and $s$. This shows that $K_E(U)\ge\log_2 9 - 16/9 \approx 1.392$ ebits for all $U$ in case (I.1) .

The second case (I.2) is that $p>0$ and $V_2\ne V_4$. Then both of $V_2$ and $V_4$ are nonzero. Up to local permutation matrices on $\cH_B$, we may assume
\bea
\label{eq:v2v4=}
V_2&=&I_s\op [V_{21},V_{22}],\notag\\
V_4&=&[V_{41},V_{42}]\op I_t
\eea
with $s,t\ge0$, where the submatrices $V_{21}$ and $V_{42}$ act on the same subspace $\lin\{\ket{s+1},\cdots,\ket{p-t}\}$ of dimension $p-s-t$.
The moves in the paragraph including \eqref{eq:ud1} imply that $p>s+t$. So $V_{21}$ and $V_{42}$ are both nonzero, and are in the column vectors of the same position of $V_2$ and $V_4$. Furthermore $V_{21}$ and $V_{42}$ respectively have no nonzero diagonal entries of $V_2$ and $V_4$. So $\left(
                   \begin{array}{cc}
                     0 \\
                     V_{21}
                     \end{array}
                 \right)\ket{j}\ne\ket{j}$ and $\left(
                   \begin{array}{cc}
                     V_{42} \\
                     0
                     \end{array}
                 \right)\ket{j}\ne\ket{j}$ for all $j\in[s+1,p-t]$. Note that $V_{21}$ and $V_{42}$ are both of full rank. If
$\left(
                   \begin{array}{cc}
                     0 \\
                     V_{21}
                     \end{array}
                 \right)\ket{j}=\left(
                   \begin{array}{cc}
                     V_{42}  \\
                     0
                     \end{array}
                 \right)\ket{j}$ for all $j\in[s+1,p-t]$, then $V_{21}=\left(
                   \begin{array}{cc}
                     X  \\
                     0 \\
                     \end{array}
                 \right)$ and $V_{42}=\left(
                   \begin{array}{cc}
                     0  \\
                     X \\
                     \end{array}
                 \right)$
with a permutation matrix $X$, and thus from \eqref{eq:v2v4=} we obtain that $V_2$ and $V_4$ are both equal to $X$ up to the moves in the paragraph including \eqref{eq:ud1}. This is a contradiction with the assumption at the beginning of this paragraph.
Hence, we can find out some $j\in[s+1,p-t]$, such that $\left(
                   \begin{array}{cc}
                     0 \\
                     V_{21}
                     \end{array}
                 \right)\ket{j}\ne \left(
                   \begin{array}{cc}
                     V_{42}
                     \\
                     0
                     \end{array}
                 \right)\ket{j}$. It implies that $\ket{j}$, $V_2\ket{j}$ and
$V_4\ket{j}$ are pairwise orthogonal. Let $U$ act on the product state
$
{1\over\sqrt3}
( \ket{a_1}
+\ket{a_2}
+\ket{a_3}) \ket{j},
$
where the state $\ket{a_j}$ satisfies $D_j\ket{a_k}=\d_{jk}\ket{a_j}$, i.e., $D_j$ is the stabilizer of $\ket{a_j}$. So the resulting state ${1\over\sqrt3}(\ket{a_1}\ket{j}+\ket{a_2}V_2\ket{j}+\ket{a_3}V_4\ket{j})$ is a Schmidt-rank-three maximally entangled state, and we have created $\log_2 3$ ebits.

The third case (I.3) is that $p>0$ and $V_2=V_4$. So we may assume that $V_2$ does not have nonzero diagonal entries, and thus $p>1$. Since $U$ has Schmidt rank three, $n$ and $q$ are not simultaneously zero. If $n=0$ or $q=0$, by performing the local permutation matrix $I_A\ox (I_{m+n+q}\op V_2^\dg)$ on the lhs of $U$, we obtain a new unitary of the type of case (I.1). Thus we may assume $n>0$ and $q>0$. Since $V_1$ and $V_3$ have no nonzero diagonal entries, we have $n>1$ and $q>1$.
Since the identity matrix and any permutation matrix are simultaneously diagonalizable, $U$ is locally equivalent to a Schmidt-rank-three diagonal unitary.
The unitary $U$ with $m=0$ and $n=q=p=2$ can generate exactly $\log_2 3$ ebits of entanglement starting from a product state without ancillas. An optimal choice of the input state is $\frac{1}{\sqrt{3}}(1,1,1)\otimes (g,h,g,h,g,h)$, where $g=\frac{1+\sqrt{3}}{2\sqrt{6}}$, $h=\frac{1-\sqrt{3}}{2\sqrt{6}}$.
For generic cases in the case (I.3), we may assume $m=0$ for the same reason as in case (I.1) above, and consider $n,q,p$ to be integers not less than two. Up to local permutation unitaries there is a cyclic shift (of length $t,s,u$ respectively) in each of the three permutation unitaries $V_1$, $V_2$ and $V_3$, and we let the input state to have nonzero amplitude on the support of these operators only and let them of the form $\frac{1}{\sqrt{3}}(1,1,1)\otimes (v_1,v_2,v_3)$, where the $v_1,v_2,v_3$ are real vectors of length $t,s,u$, respectively. The elements in $v_1$ are just two real numbers appearing alternately: $e,f,e,f,\dots$, and thus the last number in $v_1$ is $e$ if $t$ is odd, and is $f$ if $t$ is even. Similarly, $v_2=(g,h,g,h,\dots)$, and the last number in $v_2$ is $g$ if $s$ is odd, and is $h$ if $s$ is even. And $v_3=(y,z,y,z,\dots)$, and the last number in $v_3$ is $y$ if $u$ is odd, and is $z$ if $u$ is even. Then the maximal output entanglement of $\log_2 3$ ebits is achievable, by choosing $e,f,g,h,y,z\in\mathbb{R}$ which satisfy that $e-f=\frac{1}{\sqrt{2\lfloor t/2\rfloor}}$, $g-h=\frac{1}{\sqrt{2\lfloor s/2\rfloor}}$, $y-z=\frac{1}{\sqrt{2\lfloor u/2\rfloor}}$, and $\vert v_1\vert=\vert v_2\vert=\vert v_3\vert=\frac{1}{\sqrt{3}}$. It is not hard to see that there are $2^3=8$ real solutions $(e,f,g,h,y,z)$ to the equations above, for any $t,s,u$. And since $K_E(U)\le\log_2 r$ ebits for any $U$ of Schmidt rank $r$, we have that $K_E(U)=\log_2 3$ ebits for all $U$ in case (I.3).

Case (II). Suppose $U$ is a Schmidt-rank-three controlled permutation unitary with four terms, and is controlled from the $A$ side.
By following similar arguments as in (I) but also noting that the $B$-side operators in all four terms in $U$ are permutation matrices, it can be shown that up to local permutation unitaries, the $U$ is of the form
\bea\label{eq:perm_u_4terms}
U=
& D_1\ox I_B + D_2 \ox (I_m \op I_n \op V_1) +
\notag\\
& D_3 \ox (I_m \op V_2 \op I_q)+ D_4 \ox (I_m \op V_2 \op V_1),\quad
\eea
where $D_j$ ($j=1,\dots,4$) are orthogonal projectors onto the computational basis states that add up to $I_A$, while $V_1$ and $V_2$ are permutation matrices of size $q\times q$ and $n\times n$, respectively, and their diagonal elements are all zero. And $m\ge 0$ is an integer. Again, for the purpose of studying the entangling power of $U$, we may assume that all $D_j$ in \eqref{eq:perm_u_4terms} are one-dimensional projectors.

When $q=n=2$, the entangling power of $U$ is exactly $\log_2 3$ ebits, and this number is achieved by a product input state without ancillas. For example, when $m=0$, there is an input state of the form $\frac{1}{2}(1,1,1,1)\otimes (g,h,g,h)$ which gives the optimal output entanglement, where $g=\frac{\sqrt{3}+\sqrt{6}}{6}$, and $h=\frac{\sqrt{3}-\sqrt{6}}{6}$ are the same numbers as in case (I.1). When $m>0$, we choose the $B$-side input state so that it has zero amplitude in the support of $I_m$ in \eqref{eq:perm_u_4terms}, then we are back to the $m=0$ case. For other values of $q$ and $n$, and arbitrary $m\ge 0$ (which is treated as $m=0$), we also have that the entangling power of $U$ is exactly $\log_2 3$ ebits. A class of the optimal input states is the same as those in case (I.1), although it is possible that there are other classes of optimal input states as well.

Case (III). Now the only remaining case is that $U$ is of the form of the last case in Lemma~\ref{lm:sch3perm}(i).  An example of this case is in \eqref{eq:uketbra11}. When no ancillas are allowed, the $U$ in \eqref{eq:uketbra11} can generate at most 1 ebit, since it is on a $3\times 2$ dimensional system. When ancillas are allowed, we choose the ancillas $A'$ and $B'$ to be of the same size as the input systems $A$ and $B$, respectively, and let the input state on the two sides be the maximally entangled states $\sum_{j=1}^3 \ket{jj}_{AA'}$ and $\sum_{k=1}^2 \ket{kk}_{BB'}$, respectively, then the output state contains exactly $\log_2 3$ ebits. For other unitaries $U$ of the type of case (III), up to local permutations and a swap of the two systems we may write $U$ as
\bea
\label{eq:uketbra11b}
U&=&(P_A\ox V_B)
\notag\\
&\oplus_A&
[(I_A-P_A)\ox Q_B + W_A \ox (I_B-Q_B)],
\eea
where $P_A$ and $Q_B$ are projectors onto computational basis states of $\cH_A$ and $\cH_B$, respectively, and $W_A$ is a partial permutation matrix which is of full rank in the support of $I_A-P_A$, and $V_B$ is a permutation matrix. We choose the input state on the $A$ side to be of the form $\sum_{j=1}^{d_A} \mu_j\ket{jj}$, where the real coefficients $\mu_j$ take at most three different values including zero, and $\mu_j=0$ iff $\bra{j}W_A\ket{j}\ne 0$. The nonzero values of $\mu_j$ are the same for $\ket{j}_A$ in the support of $P_A$. And the same statement holds for the support of $I_A-P_A$.
And choose the input state on the $B$ side to be $\sum_{k=1}^{d_B} \nu_k \ket{kk}_{BB'}$, where the real coefficients $\nu_k$ take at most three different values including zero, and $\nu_k=0$ iff $\bra{k}V_B\ket{k}\ne 0$. The nonzero values of $\nu_k$ are the same for $\ket{k}_B$ in the support of $Q_B$. And the same statement holds for the support of $I_B-Q_B$. With a suitable choice of the $\mu_j$ and $\nu_k$ subject to the constraints above, the output entanglement is exactly $\log_2 3$ ebits. Therefore, the entangling power of $U$ in case (III) is always $\log_2 3$ ebits.

In summary, we have considered all forms of $U$, and thus the assertion holds.
\epf

\bibliographystyle{unsrt}

\bibliography{channelcontrol}

\end{document}